\documentstyle[11pt,epsfig]{article}
\def\halpha{H$\alpha$}
\def\sfr{$M_\odot$~yr$^{-1}$}
\def\msun{$M_\odot$}
\def\microns{$\mu$m}

\begin{document}

{\large{\bf STAR FORMATION IN GALAXIES ALONG THE HUBBLE SEQUENCE}} 

\bigskip

{\large{\it Robert C. Kennicutt, Jr}}

Steward Observatory, The University of Arizona \\
Tucson, Arizona  85721 \\
rkennicutt@as.arizona.edu

\bigskip

\centerline{ABSTRACT}

\medskip

Observations of star formation rates (SFRs) in galaxies 
provide vital clues to the physical nature of the Hubble sequence, and  
are key probes of the evolutionary histories of galaxies.  The focus
of this review is on the broad patterns in the star formation properties
of galaxies along the Hubble sequence, and their implications for 
understanding galaxy evolution and the physical processes that drive the 
evolution.  Star formation in the disks and nuclear regions of galaxies are 
reviewed separately, then discussed within a common interpretive framework.
The diagnostic methods used to measure SFRs are also reviewed, and
a self-consistent set of SFR calibrations is presented as a aid
to workers in the field.

\medskip


\bigskip

To appear in Vol.~36 of {\it Annual Review of Astronomy and Astrophysics}

\bigskip

\section{INTRODUCTION}

One of the most recognizable features of galaxies along the Hubble sequence
is the wide range in young stellar content and star formation activity.
This variation in stellar content is part of the basis of 
the Hubble classification
itself (Hubble 1926), and understanding its physical nature and
origins is fundamental to understanding galaxy evolution in its broader context.
This review deals with the global star formation properties of galaxies,
the systematics of those properties along the Hubble sequence, and their
implications for galactic evolution.  I interpret
``Hubble sequence" in this context very loosely, to encompass not only
morphological type but other properties such as gas content, mass,
bar structure, and dynamical environment, which can strongly influence
the large-scale star formation rate (SFR).

Systematic investigations of the young stellar content of galaxies trace back
to the early studies of resolved stellar populations by Hubble and Baade, 
and analyses of galaxy colors and spectra by Stebbins, Whitford, Holmberg,
Humason, Mayall, Sandage, Morgan, and de Vaucouleurs (see Whitford 1975 for
a summary of the early work in this field).  This piecemeal information 
was synthesized by Roberts (1963), in an article for the first volume
of the {\it Annual Review of Astronomy and Astrophysics}.  
Despite the limited information that was available on the 
SFRs and gas contents of galaxies, Roberts' analysis established the basic  
elements of the contemporary picture of the Hubble sequence as a monotonic
sequence in present-day SFRs and past star formation histories.

Quantifying this picture required the development of more precise 
diagnostics of global SFRs in galaxies.  The first quantitative SFRs
were derived from evolutionary synthesis models of galaxy colors
(Tinsley 1968, 1972, Searle et al 1973).  These
studies confirmed the trends in SFRs and star formation histories along
the Hubble sequence, and led to the first predictions of the evolution of the
SFR with cosmic lookback time.  Subsequent modelling of blue 
galaxies by Bagnuolo (1976), Huchra (1977), and Larson \& Tinsley (1978) 
revealed the importance of star formation bursts in the evolution of low-mass 
galaxies and interacting systems.  Over the next decade the field matured
fully, with the development of more precise direct SFR diagnostics,
including integrated emission-line fluxes (Cohen 1976, Kennicutt 1983a), 
near-ultraviolet continuum fluxes (Donas \& Deharveng 1984), 
and infrared continuum fluxes (Harper \& Low 1973, Rieke \& Lebofsky 1978,
Telesco \& Harper 1980).  These provided 
absolute SFRs for large samples of nearby galaxies, and these were 
subsequently interpreted in terms of the evolutionary properties of  
galaxies by Kennicutt (1983a), Gallagher et al (1984), and Sandage (1986).

Activity in this field has grown enormously in the past
decade, stimulated in large part by two major revelations.  The 
first was the discovery of a large population of ultraluminous infrared 
starburst galaxies by the Infrared Astronomical Satellite (IRAS) in 
the mid-1980's.  Starbursts had been identified (and coined) from   
groundbased studies (Rieke \& Lebofsky 1979; Weedman et al 1981), 
but IRAS revealed 
the ubiquity of the phenomenon and the extreme nature of the most luminous
objects.  The latest surge of interest in the field
has been stimulated by the detection of star forming galaxies at high
redshift, now exceeding $z=3$ (Steidel et al 1996, Ellis 1997).  
This makes it possible to apply the locally calibrated SFR 
diagnostics to distant galaxies, and directly trace the evolution of 
the SFR density and the Hubble sequence with cosmological lookback time.

The focus of this review is on the broad patterns in the star formation
properties of galaxies, and their implications for the evolutionary
properties of the Hubble sequence.
It begins with a summary of the diagnostic methods used to measure 
SFRs in galaxies, followed by a summary of the systematics of SFRs 
along the Hubble sequence, and the interpretation of those trends
in terms of galaxy evolution.  It concludes with a brief discussion 
of the physical regulation of the SFR in galaxies and future prospects
in this field.  Galaxies exhibit a huge dynamic range in 
SFRs, over six orders of magnitude even when normalized per unit area
and galaxy mass, and the continuity of physical properties over this 
entire spectrum of activities is a central theme of this review. 

With this broad approach in mind, I cannot begin to review the hundreds of
important papers on the star formation properties of individual galaxies,
or the rich theoretical literature on this subject.  Fortunately, there
are several previous reviews in this series that provide thorough 
discussions of key aspects of this field.  A broad review of the physical
properties of galaxies along the Hubble sequence can be found in
Roberts \& Haynes (1994).  The star formation and evolutionary properties
of irregular galaxies are reviewed by Gallagher \& Hunter (1984).
The properties of IR-luminous starbursts are the subject of several 
reviews, most recently those by Soifer et al (1987), Telesco 
(1988), and Sanders \& Mirabel (1996).  Finally an excellent review
of faint blue galaxies by Ellis (1997) describes many applications
to high-redshift objects.

\section{DIAGNOSTIC METHODS}

Individual young stars are unresolved in all but the closest 
galaxies, even with the {\it Hubble Space Telescope} (HST),
so most information on the star formation properties of galaxies
comes from integrated light measurements in the ultraviolet (UV), 
far-infrared (FIR),
or nebular recombination lines.  These direct tracers of the young
stellar population have largely supplanted earlier SFR measures based 
on synthesis modelling of broadband colors, though the latter are still
applied to multicolor observations of faint galaxies.  This section 
begins with a brief discussion of synthesis models, which form the basis
of all of the methods, followed by more detailed discussions of the
direct SFR tracers.

\subsection{{\it Integrated Colors and Spectra, Synthesis Modelling}}

The basic trends in galaxy spectra with Hubble type are illustrated
in Figure 1, which shows examples of integrated spectra for  
E, Sa, Sc, and Magellanic irregular galaxies (Kennicutt 1992b).
When progressing along this sequence, several 
changes in the spectrum are apparent: a broad rise in
the blue continuum, a gradual change in the composite stellar 
absorption spectrum from K-giant dominated to A-star dominated, and 
a dramatic increase in the strengths of the nebular emission lines,
especially \halpha.

Although the integrated spectra contain contributions from the full
range of stellar spectral types and luminosities, it is easy to
show that the dominant contributors at visible wavelengths are 
intermediate-type main sequence stars (A to early F) and 
G-K giants.  As a result, the integrated colors and spectra of 
normal galaxies fall on a relatively tight sequence, with the
spectrum of any given object dictated by the ratio of early to
late-type stars, or alternatively by the ratio of young ($< 1$~Gyr) to
old (3--15~Gyr) stars.  This makes it possible to use the observed
colors to estimate the fraction of young stars and 
the mean SFR over the past $10^8$--$10^9$ years.

\begin{figure}[!ht]
  \begin{center}
    \leavevmode
  \centerline{\epsfig{file=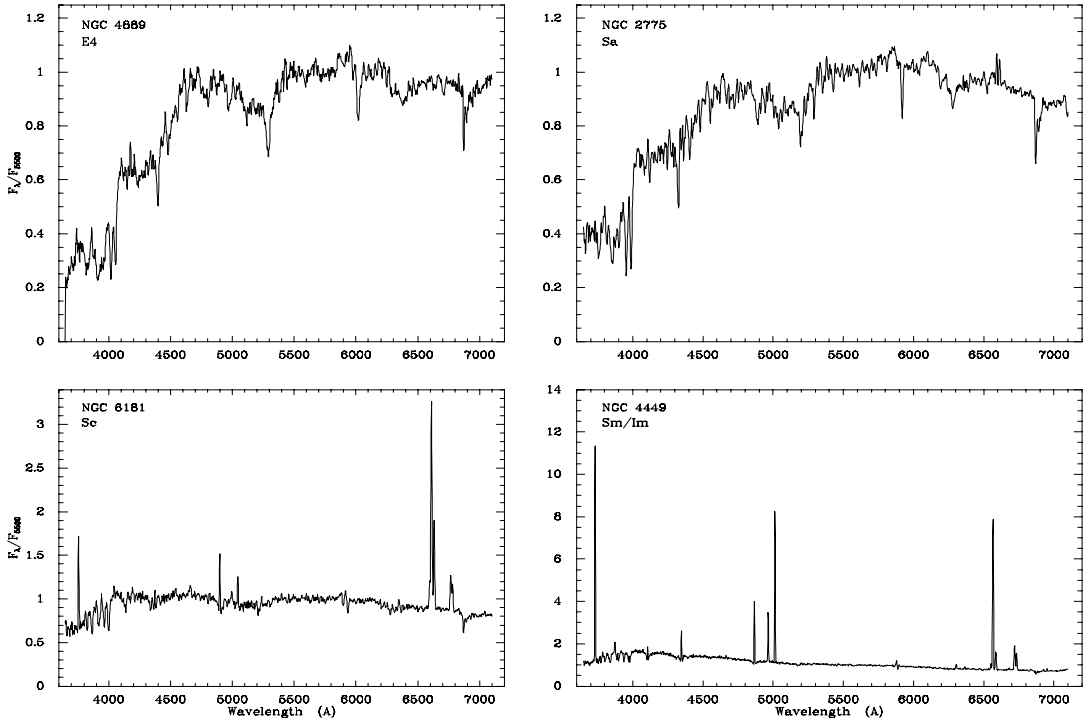,width=16cm}}
  \end{center}
  \caption{\em Integrated spectra of elliptical, spiral, and irregular 
galaxies, from Kennicutt (1992b).  The fluxes have been normalized to 
unity at 5500~\AA.}
\end{figure}

The simplest application of this method would assume a linear scaling 
between the SFR and the continuum luminosity integrated over a fixed
bandpass in the blue or near-ultraviolet.  Although 
this may be a valid approximation in starburst galaxies, where young
stars dominate the integrated light across the visible spectrum, the
approximation breaks down in most normal 
galaxies, where a considerable fraction of the continuum is produced by
old stars, even in the blue (Figure 1).  However the scaling of the SFR
to continuum luminosity is a smooth function of the color of the population,
and this can be calibrated using an evolutionary synthesis model.

Synthesis models are used in all of the methods described here, so it is
useful to summarize the main steps in the construction of a model. 
A grid of stellar evolution tracks is used to derive
the effective temperatures and bolometric luminosities for various
stellar masses as a function of time, and these are converted into
broadband luminosities (or spectra) using stellar atmosphere models or 
spectral libraries.  The individual stellar templates are then summed together, 
weighted by an initial mass function (IMF), to synthesize the luminosities, 
colors, or spectra of single-age populations as functions of age.
These isochrones can then be added
in linear combination to synthesize the spectrum or colors of a galaxy
with an arbitrary star formation history, usually parametrized as
an exponential function of time.  Although a single model contains at
least four free parameters, the star formation history, galaxy age, metal   
abundance, and IMF, the colors of normal galaxies are well represented by
a one-parameter sequence with fixed age, composition and IMF, varying only
in the time dependence of the SFR (Searle et al 1973, Larson \&
Tinsley 1978; Charlot \& Bruzual 1991).  

Synthesis models have been published by several authors, and are often
available in digital form.  An extensive library of models has been compiled
by Leitherer et al (1996a), and the models are described in a 
companion conference volume (Leitherer et al 1996b).  Widely used  
models for star forming galaxies include those of Bruzual \& Charlot
(1993), Bertelli et al (1994), and Fioc \& Rocca-Volmerange (1997).
Leitherer \& Heckman (1995) have published an extensive grid of models that 
is optimized for applications to starburst galaxies.

\begin{figure}[!ht]
  \begin{center}
    \leavevmode
  \centerline{\epsfig{file=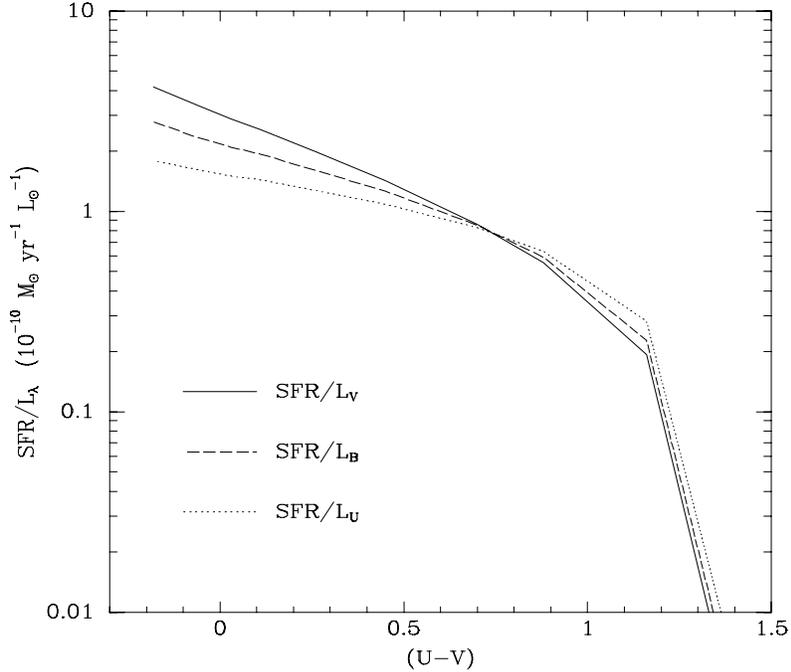,width=16cm}}
  \end{center}
  \caption{\em  Relationship between SFR per unit broadband luminosity in
the $UBV$ passbands and integrated color, from the evolutionary synthesis
models of Kennicutt et al (1994).  The models are for 10-billion-year-old disks,
a Salpeter IMF, and exponential star formation histories.  The $U$, $B$,
and $V$ luminosities are normalized to those of the Sun in the respective
bandpasses.}
\end{figure}

The synthesis models provide relations between the SFR per unit mass 
or luminosity and the integrated color of the population.  An example is
given in Figure 2, which plots the SFR per unit $U$, $B$, and $V$ 
luminosity as functions of $U-V$ color, based on the models of
Kennicutt et al (1994).  
Figure 2 confirms that the broadband luminosity by itself is a poor
tracer of the SFR; even the SFR/$L_U$ ratio varies
by more than an order of magnitude over the relevant range of galaxy colors.
However the integrated color provides a reasonable estimate of the
SFR per unit luminosity, especially for the bluer galaxies.

SFRs derived in this way are relatively imprecise, and are prone to 
systematic errors from reddening or from an incorrect IMF, age, metallicity,
of star formation history (Larson \& Tinsley 1978).  Nevertheless, the 
method offers a useful means 
of comparing the average SFR properties of large samples of galaxies, when 
absolute accuracy is not required.  The method should be avoided in 
applications where the dust content, abundances, or IMFs are likely to 
change systematically across a population.

\subsection{{\it Ultraviolet Continuum}}

The limitations described above 
can be avoided if observations are made at wavelengths where the integrated 
spectrum is dominated by young stars, so that the SFR scales
linearly with luminosity.  The optimal wavelength range is 
1250--2500~\AA, longward of the Ly$\alpha$ forest but short
enough to minimize spectral contamination from older stellar
populations.  These wavelengths are inaccessible from the ground for
local galaxies ($z<0.5$), but the region can be observed in the redshifted
spectra of galaxies at $z\sim$1--5.  The recent detection of the redshifted
UV continua of large numbers of $z>3$ galaxies with the Keck
telescope has demonstrated the enormous potential of this technique
(Steidel et al 1996).

The most complete UV studies of nearby galaxies are based on dedicated   
balloon, rocket, and space experiments  
(Smith \& Cornett 1982, Donas \& Deharveng 1984, Donas et al
1987, 1995, Buat 1992, Deharveng et al 1994).  The database of 
high-resolution UV imaging of galaxies is improving rapidly, mainly
from HST (Meurer et al 1995, Maoz 1996) and the {\it Ultraviolet Imaging 
Telescope} (Smith et al 1996, Fanelli et al 1997).  An atlas of 
UV spectra of galaxies from the {\it International Ultraviolet Explorer} 
has been published by Kinney et al (1993).  A recent conference volume
by Waller et al (1997) highlights recent UV observations of galaxies.

The conversion between the UV flux over a given wavelength interval and 
the SFR can be derived using the synthesis models described earlier.
Calibrations have been published by Buat et al (1989), Deharveng et al 
(1994), Leitherer et al (1995b), Meurer et al (1995), Cowie et al (1997), 
and Madau et al (1998), for wavelengths in the range 1500--2800~\AA.
The calibrations differ over a full range of $\sim$0.3 dex, when converted 
to a common reference wavelength and IMF, with most of the difference  
reflecting the use of different stellar libraries or different assumptions
about the star formation timescale.  For integrated measurements of galaxies,
it is usually appropriate to assume that the SFR has remained constant
over timescales that are long compared to the lifetimes of the dominant
UV emitting population ($<$10$^8$ yr), in the ``continuous star formation"
approximation.  Converting the calibration of Madau et al (1998) to a 
Salpeter (1955) IMF with mass limits 0.1 and 100 M$_\odot$ yields:
\begin{equation}
{\rm SFR}~(M_\odot~yr^{-1}) = 1.4 \times 10^{-28}~L_\nu~({\rm ergs~s^{-1}~Hz^{-1}}).
\end{equation}

For this IMF, the composite UV spectrum happens to be 
nearly flat in $L_\nu$, 
over the wavelength range 1500--2800~\AA, and this allows us to express
the conversion in Equation 1 in such simple form.  
The corresponding conversion in terms of $L_\lambda$ will 
scale as $\lambda^{-2}$.  Equation 1 applies to galaxies with continuous
star formation over timescales of $10^8$ years or longer; the SFR/$L\nu$
ratio will be significantly lower in younger populations such as young
starburst galaxies.  For example, continuous burst models for a 9 Myr old
population yield SFRs that are 57\%\ higher than those given in Equation 1 
(Leitherer et al 1995b).  It is important when using this method 
to apply an SFR calibration that is appropriate to the population of interest.

The main advantages of this technique are that it is directly tied to the
photospheric emission of the young stellar population, and it can be
applied to star forming galaxies over a wide range of redshifts.  
As a result, it is currently the most powerful probe of the 
cosmological evolution in the SFR (Madau et al 1996, Ellis 1997).
The chief drawbacks of the method are its sensitivity to extinction and
the form of the IMF.  Typical extinction corrections in the integrated
UV magnitudes are 0--3 magnitudes (Buat 1992, Buat \& Xu 1996).  
The spatial distribution of the extinction is very patchy, with the 
emergent UV emission being dominated by regions of relatively low 
obscuration (Calzetti et al 1994), so calibrating the extinction
correction is problematic.  The best determinations
are based on two-component radiative transfer models which 
take into account the clumpy distribution of dust, and make use 
of reddening information from the Balmer decrement or IR
recombination lines (e.g., Buat 1992,
Calzetti et al 1994, Buat \& Xu 1996, Calzetti 1997).  

The other main limitation, which is shared by all of the
direct methods, is the dependence of the derived SFRs on the 
assumed form of the IMF.  The integrated spectrum in the 1500--2500~\AA\
range is dominated by stars with masses above $\sim$5~\msun, so the
SFR determination involves a large extrapolation to lower stellar masses.
Fortunately there is little evidence for large systematic variations
in the IMF among star forming galaxies (Scalo 1986, Gilmore et al 1998), with
the possible exception of IR-luminous starbursts, where the UV
emission is of little use anyway.

\subsection{{\it Recombination Lines}}

Figure 1 shows that the most dramatic change
in the integrated spectrum with galaxy type is a rapid increase in
the strengths of the nebular emission lines.  The nebular lines effectively 
re-emit the integrated stellar luminosity of galaxies shortward of the 
Lyman limit, so they provide a direct, sensitive probe of the young massive
stellar population.  Most applications of this method have been based on 
measurements of the \halpha\ line, but 
other recombination lines including H$\beta$, P$\alpha$, P$\beta$, 
Br$\alpha$, and Br$\gamma$ have been used as well.

The conversion factor between ionizing flux and the SFR is usually computed
using an evolutionary synthesis model.  Only stars
with masses $>$10~\msun\ and lifetimes $<$20 Myr
contribute significantly to the integrated ionizing flux, 
so the emission lines provide a nearly instantaneous measure of the 
SFR, independent of the previous star formation history.  Calibrations have 
been published by numerous authors, including Kennicutt (1983a), 
Gallagher et al (1984), Kennicutt et al (1994), Leitherer \& Heckman (1995),
and Madau et al (1998).  For solar abundances and the same Salpeter IMF 
(0.1--100~\msun) as was used in deriving equation [1], the calibrations 
of Kennicutt et al (1994) and Madau et al (1998) yield:
\begin{equation}
{\rm SFR}~(M_\odot~yr^{-1}) = 7.9 \times 10^{-42}~L(H\alpha)~({\rm ergs~s^{-1}})
 = 1.08 \times 10^{-53}~Q(H^0)~({\rm s^{-1}}). 
\end{equation}

where $Q(H^0)$ is the ionizing photon luminosity, and the H$\alpha$
calibration is computed for Case B recombination at $T_e$ = 10000~K.  
The corresponding 
conversion factor for $L$(Br$\gamma$) is $8.2 \times 10^{-40}$ in the
same units, and it is straightforward to derive conversions for other 
recombination lines.  Equation 2 yields SFRs that are 
7\%\ lower than the widely used calibration of Kennicutt (1983a), with
the difference reflecting a combination of updated stellar models
and a slightly different IMF (Kennicutt et al 1994).  As with other methods,
there is a significant variation among published calibrations ($\sim$30\%),
with most of the dispersion reflecting differences in the stellar
evolution and atmosphere models.  

Large \halpha\ surveys
of normal galaxies have been published by Cohen (1976), Kennicutt \& 
Kent (1983), Romanishin (1990), Gavazzi et al (1991), 
Ryder \& Dopita (1994), Gallego et al (1995), and Young et al (1996).
Imaging surveys have been published by numerous other authors, with some
the largest including Hodge \& Kennicutt (1983), Hunter \& Gallagher (1985), 
Ryder \& Dopita (1993), Phillips (1993), Evans et al (1996), 
Gonz\'alez Delgado et al (1997), and Feinstein (1997).  Gallego et 
al (1995) have observed a complete emission-line selected sample, in order
to measure the volume-averaged SFR in the local universe, and this
work has been applied extensively to studies of the evolution in the 
SFR density of the universe (Madau et al 1996).

The primary advantages of this method are its high sensitivity, and 
the direct coupling between the nebular emission and the massive SFR.
The star formation in nearby galaxies can be mapped at high resolution 
even with small telescopes, and the \halpha\ line can be detected in
the redshifted spectra of starburst galaxies to $z$$\gg$2 (e.g. Bechtold
et al 1997).
The chief limitations of the method are its sensitivity to uncertainties
in extinction and the IMF, and to the assumption that all of the
massive star formation is traced by the ionized gas.  The escape fraction
of ionizing radiation from individual HII regions has been measured 
both directly (Oey \& Kennicutt 1997) and from observations of the diffuse 
\halpha\ emission in nearby galaxies (e.g., Hunter et al 1993, Walterbos \& 
Braun 1994, Kennicutt et al 1995, Ferguson et al 1996, Martin 1997), with
fractions of 15--50\%\ derived in both sets of studies.  Thus it is 
important when using this method to include the diffuse \halpha\ emission 
in the SFR measurement (Ferguson et al 1996).  However the escape fraction
from a galaxy as a whole should be much lower.  Leitherer et al (1995a) 
directly measured the redshifted Lyman continuum region in four starburst 
galaxies, and they derived an upper limit of 3\%\ on the escape fraction of 
ionizing photons.  Much higher global 
escape fractions of 50--94\%, and local escape fractions as high as 
99\%\ have been estimated by Patel \& Wilson (1995a, b), based on
a comparison of O-star densities and \halpha\ luminosities in M33 and 
NGC~6822, but those results are subject to large uncertainties,
because the O-star properties and SFRs were derived from
$UBV$ photometry, without spectroscopic identifications.
If the direct limit of $<$3\%\ from Leitherer et al (1995a) 
is representative, then density bounding effects are a negligible source
of error in this method.  However it is very important to test this
conclusion by extending these types of measurements to a more diverse 
sample of galaxies.  

Extinction is probably the most important source of systematic error in 
\halpha-derived SFRs.  The extinction can be 
measured by comparing \halpha\ fluxes with those of IR
recombination lines or the thermal radio continuum.  Kennicutt (1983a) and   
Niklas et al (1997) have used integrated \halpha\ and radio fluxes of galaxies
to derive a mean extinction $A$(\halpha) = 0.8--1.1 mag.  Studies of large 
samples of individual HII regions in nearby galaxies yield similar results, 
with mean $A$(\halpha) = 0.5--1.8 mag (e.g. Caplan \& Deharveng
1986, Kaufman et al 1987, van der Hulst et al 1988, Caplan et al 1996).

Much higher extinction is encountered in localized regions, especially in the
the dense HII regions in circumnuclear starbursts, and there the near-IR 
Paschen or Brackett recombination lines are required to reliably 
measure the SFR.  Compilations of these data include Puxley et al (1990),
Ho et al (1990), Calzetti et al (1996), Goldader et al (1995, 1997), 
Engelbracht (1997), and references therein.  The Paschen and Brackett lines 
are typically 1--2 orders of magnitude weaker than \halpha, so most 
measurements to date have been restricted to high surface brightness 
nuclear HII regions, but it is gradually becoming feasible 
to extend this approach to galaxies as a whole.  The same method can
be applied to higher-order recombination lines or the thermal continuum
emission at submillimeter and radio wavelengths.  Examples of such 
applications include H53$\alpha$ measurements of M82
by Puxley et al (1989), and radio continuum measurements of disk galaxies
and starbursts by Israel \& van der Hulst (1983), Klein \& Grave (1986),
Turner \& Ho (1994), and Niklas et al (1995).

The ionizing flux is produced almost exclusively by stars with
M $>$ 10~M$_\odot$, so SFRs derived from this method are especially sensitive
to the form of the IMF.   Adopting the Scalo (1986) IMF, for example, yields 
SFRs that are $\sim$3 times higher than derived with a Salpeter IMF.
Fortunately, the \halpha\ equivalent widths and
broadband colors of galaxies are very sensitive to the slope of the 
IMF over the mass range 1--30~M$_\odot$, and these can be used to
constrain the IMF slope (Kennicutt 1983a, Kennicutt et al 1994).  
The properties of normal disks are well fitted by a Salpeter IMF
(or by a Scalo function with Salpeter slope above 1~M$_\odot$),
consistent with observations of resolved stellar populations in
nearby galaxies (e.g. Massey 1998).  As with the UV continuum method,
it is important when applying published SFRs to take proper account of
the IMF that was assumed.

\subsection{{\it Forbidden Lines}}

The \halpha\ emission line is redshifted out of the visible window beyond 
$z$$\sim$0.5, so there is considerable interest in calibrating bluer 
emission lines as quantitative SFR tracers.  Unfortunately the integrated
strengths of H$\beta$ and the higher order Balmer emission lines are poor
SFR diagnostics, because the lines are weak and stellar absorption
more strongly influences the emission-line fluxes.  These lines in fact are
rarely seen in emission at all in the integrated spectra of galaxies
earlier than Sc (Kennicutt 1992a, also see Figure 1).

The strongest emission feature in the blue is the [OII]$\lambda$3727
forbidden-line doublet.  The luminosities of forbidden lines are not
directly coupled to the ionizing luminosity, and their excitation is
sensitive to abundance and the ionization state of the gas.  However
the excitation of [OII] is sufficiently well behaved that it can be
calibrated empirically (through \halpha) as a quantitative SFR tracer.
Even this indirect calibration is extremely useful for 
lookback studies of distant galaxies, because [OII] can be observed
in the visible out to redshifts $z \sim 1.6$, and it has been measured
in several large samples of faint galaxies  (Cowie et al 1996, 1997,
Ellis 1997, and references therein).

Calibrations of SFRs in terms of [OII] luminosity have been published 
by Gallagher et al (1989), based on large-aperture
spectrophotometry of 75 blue irregular galaxies, and by Kennicutt (1992a), 
using integrated spectrophotometry of 90 normal and peculiar galaxies.  
When converted to the same IMF and \halpha\ calibration the resulting SFR
scales differ by a factor of 1.57, reflecting excitation differences in the
two samples.  Adopting the average of these calibrations yields:
\begin{equation}
SFR~(M_\odot~yr^{-1}) = {(1.4 \pm 0.4)} \times 10^{-41}~L[OII]~({\rm ergs~s^{-1}}),
\end{equation}

where the uncertainty indicates the range between blue emission-line
galaxies (lower limit) and samples of more luminous spiral and irregular
galaxies (upper limit).  As with Equations 1 and 2, the observed
luminosities must be corrected for extinction, in this case the
extinction at H$\alpha$, because of the manner in which the [OII] fluxes 
were calibrated.

The SFRs derived from [OII] are less precise than from \halpha, because
the mean [OII]/\halpha\ ratios in individual galaxies vary considerably,
over 0.5--1.0 dex in the Gallagher et al (1989) and Kennicutt (1992a)
samples, respectively.  The [OII]-derived SFRs may also be prone to 
systematic errors from extinction and variations in the diffuse gas 
fraction.  The excitation of [OII] is especially high in the diffuse ionized  
gas in starburst galaxies (Hunter \& Gallagher 1990, Hunter 1994, 
Martin 1997), enough to more than double the L[OII]/SFR ratio in the
integrated spectrum (Kennicutt 1992a).  On the other hand, metal 
abundance has a relatively small effect on the [OII] calibration, over
most of the abundance range of interest ($0.05~Z_\odot \le Z \le 1~Z_\odot$).
Overall the [OII] lines provide a very useful estimate of the systematics
of SFRs in samples of distant galaxies, and are especially useful as 
a consistency check on SFRs derived in other ways.

\subsection{{\it Far-Infrared Continuum}}

A significant fraction of the bolometric luminosity of a galaxy is
absorbed by interstellar dust and re-emitted in the
thermal IR, at wavelengths of roughly 10--300~$\mu$m.  The 
absorption cross section of the dust is strongly peaked in the
ultraviolet, so in principle the FIR emission can be a sensitive
tracer of the young stellar population and SFR.  The IRAS survey provides 
FIR fluxes for over
30,000 galaxies (Moshir et al 1992), offering a rich reward to those
who can calibrate an accurate SFR scale from the 10--100~$\mu$m FIR
emission.

The efficacy of the FIR luminosity as a SFR tracer depends on the
contribution of young stars to heating of the dust, 
and on the optical depth of the dust in the star forming
regions.  The simplest physical situation is one in which young stars dominate
the radiation field thoughout the UV--visible, and the dust opacity is high
everywhere, in which case the FIR luminosity measures  
the bolometric luminosity of the starburst.  In such a limiting case 
the FIR luminosity is the ultimate SFR tracer, providing
what is essentially a calorimetric measure of the SFR.  Such 
conditions roughly hold in the dense circumnuclear starbursts that
power many IR-luminous galaxies.

The physical situation is more complex in the disks of normal
galaxies, however (e.g. Lonsdale \& Helou 1987, Rowan-Robinson \& Crawford
1989, Cox \& Mezger 1989).  The FIR spectra of galaxies
contain both a ``warm" component associated with dust around young
star forming regions ($\bar{\lambda}\sim$ 60$\mu$m), and a cooler 
``infrared cirrus" component ($\bar{\lambda}\ge$ 100$\mu$m) which 
is associated with more extended dust heated by the interstellar
radiation field.  In blue galaxies, both spectral components 
may be dominated by young stars, but in red galaxies, where 
the composite stellar continuum drops off steeply in the blue, dust
heating from the visible spectra of older stars may be very important.

The relation of the global FIR emission of galaxies to the SFR 
has been a controversial subject.  In late-type
star forming galaxies, where dust heating from young stars is expected to 
dominate the 40--120$\mu$m emission, the FIR luminosity correlates with 
other SFR tracers such as the UV continuum and \halpha\ luminosities
(e.g. Lonsdale \& Helou 1987, Sauvage \& Thuan 1992, Buat \& Xu 1996).
However, early-type (S0--Sab) galaxies often exhibit high FIR
luminosities but much cooler, cirrus-dominated emission.  This emission has
usually been attributed to dust heating from the general stellar radiation 
field, including the visible radiation from older stars
(Lonsdale \& Helou 1987, Buat \& Deharveng 1988, Rowan-Robinson \& Crawford
1989, Sauvage \& Thuan 1992, 1994, Walterbos \& Greenawalt 1996).
This interpretation is supported by anomalously low
UV and \halpha\ emission (relative to the FIR luminosity) in these galaxies.
However Devereux \& Young (1990) and Devereux \& Hameed (1997) have argued
that young stars dominate the 40--120$\mu$m emission in all of these galaxies,
so that the FIR emission directly traces the SFR.  They have provided 
convincing evidence that young stars are an important source of FIR luminosity
in at least some early-type galaxies, including barred galaxies with
strong nuclear starbursts and some unusually blue objects (Section 4).
On the other hand, many early-type galaxies show
no independent evidence of high SFRs, suggesting that the older stars
or active galactic nuclei (AGNs) are responsible for much of the FIR emission.
The {\it Space Infrared Telescope Facility}, scheduled for launch early
in the next decade, should provide high-resolution FIR images of nearby
galaxies and clarify the relationship between the SFR and IR emission
in these galaxies.

The ambiguities discussed above affect the calibration of SFRs in
terms of FIR luminosity, and there probably is no single calibration
that applies to all galaxy types.  However the FIR emission should provide
an excellent measure of the SFR in dusty circumnuclear starbursts.
The SFR {\it vs} $L_{FIR}$ conversion is derived using synthesis models
as described earlier.  In the optically thick limit, it is only necessary to 
model the bolometric luminosity of the stellar population.
The greatest uncertainty in this case is the adoption of an appropriate age
for the stellar population; this may be dictated by the timescale
of the starburst itself or by the timescale for the dispersal of the
dust (so the $\tau$$\gg$1 approximation no longer holds).  
Calibrations have been published by several authors under different
assumptions about the star formation timescale (e.g. Hunter et al 1986, 
Lehnert \& Heckman 1996, Meurer et al 1997, Kennicutt 1998).  Applying
the models of Leitherer \& Heckman (1995) for continuous bursts of
age 10--100 Myr, and adopting the IMF in this paper yields the
relation (Kennicutt 1998):
\begin{equation}
{\rm SFR}~(M_\odot~yr^{-1}) = 4.5 \times 10^{-44}~L_{FIR}~({\rm ergs~s^{-1}})
~~~(starbursts),
\end{equation}

where $L_{FIR}$ refers to the infrared luminosity integrated over the
full mid and far-IR spectrum (8--1000 \microns), though for starbursts 
most of this emission
will fall in the 10--120$\mu$m region (readers should beware that 
the definition of $L_{FIR}$ varies in the literature).  Most of the other
published calibrations lie within $\pm$30\%\ of Equation 4.
Strictly speaking, the relation given above applies only to starbursts
with ages less than $10^8$ years, where the approximations applied are
valid.  In more quiescent normal star forming galaxies, the relation
will be more complicated; the contribution of dust heating from old
stars will tend to lower the effective coefficient in equation [4], whereas the
lower optical depth of the dust will tend to increase the coefficient.
In such cases, it is probably better to rely on an empirical calibration
of SFR/$L_{FIR}$, based on other methods.  For example, Buat \& Xu (1996)
derive a coefficient of $8{^{+8}_{-3}} \times 10^{-44}$, valid for
galaxies of type Sb and later only, based on 
IRAS and UV flux measurements of 152 disk galaxies.  The FIR luminosities
share the same IMF sensitivity as the other direct star formation tracers,
and it is important to be consistent when comparing results from different
sources.

\section{DISK STAR FORMATION}

The techniques described above have been used to measure SFRs in 
hundreds of nearby galaxies, and these have enabled us to  
delineate the main trends in SFRs and star formation histories along
the Hubble sequence.  Although it is customary to analyze the  
integrated SFRs of galaxies, taken as a whole, large-scale
star formation takes place in two very distinct physical environments: 
one in the extended disks of spiral and irregular galaxies, the other
in compact, dense gas disks in the centers of galaxies.  Both regimes
are significant contributors to the total star formation
in the local universe, but they are traced at different wavelengths 
and follow completely different patterns along the Hubble sequence.
Consequently I will discuss the disk and circumnuclear star formation 
properties of galaxies separately.

\subsection{{\it Global SFRs Along the Hubble Sequence}}

Comprehensive analyses of the global SFRs of galaxies have been
carried out using \halpha\ surveys (Kennicutt 1983a, Gallagher et al 1984, 
Caldwell et al 1991, 1994, Kennicutt et al 1994, Young et al 1996), 
UV continuum surveys (Donas et al 1987, Deharveng et al 1994),
FIR data (Sauvage \& Thuan 1992, Walterbos \& Greenawalt 1996, 
Tomita et al 1996, Devereux \& Hameed 1997), and multi-wavelength
surveys (Gavazzi \& Scodeggio 1996, Gavazzi et al 1996).   
The absolute SFRs in galaxies, expressed in terms of the total mass of stars 
formed per year, show an enormous range, from virtually zero in gas-poor
elliptical, S0, and dwarf galaxies to $\sim$20~M$_\odot$~yr$^{-1}$
in gas-rich spirals.  Much larger global SFRs, up to   
$\sim$100~M$_\odot$~yr$^{-1}$, can be found in optically-selected starburst 
galaxies, and SFRs as high as 1000~\sfr\ may be reached in the most luminous
IR starburst galaxies (Section 4).  The highest SFRs are associated
almost uniquely with strong tidal interactions and mergers.

\begin{figure}[!ht]
  \begin{center}
    \leavevmode
  \centerline{\epsfig{file=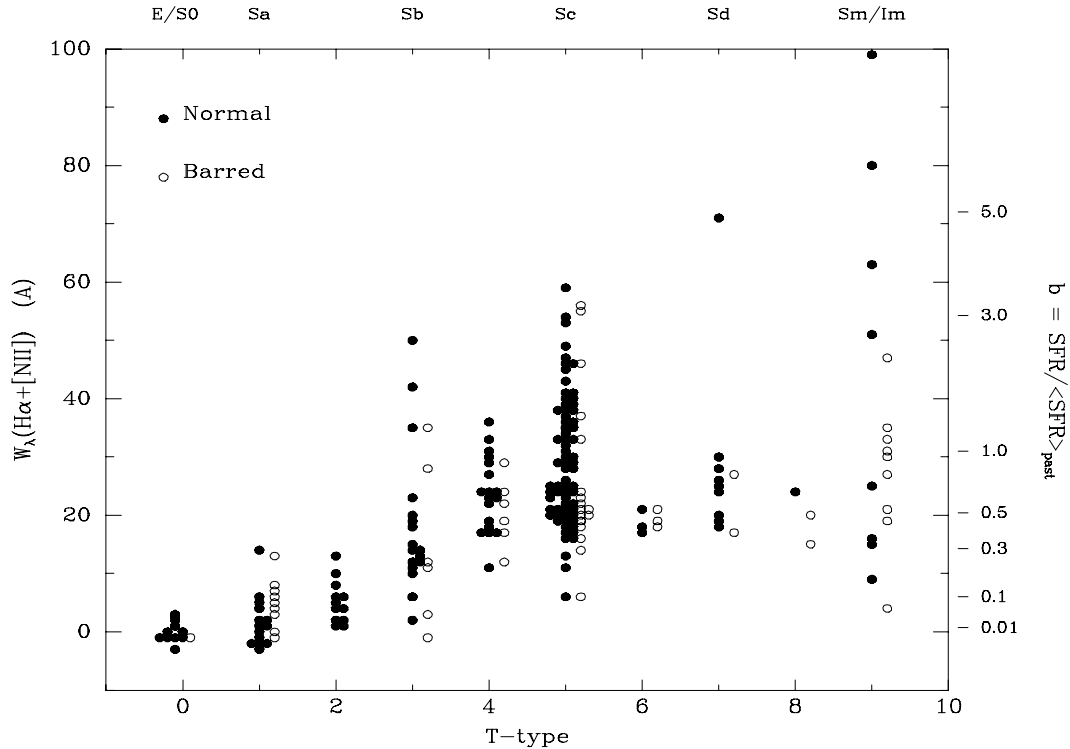,width=16cm}}
  \end{center}
  \caption{\em Distribution of integrated \halpha+[NII] emission-line 
equivalent widths for a large sample of nearby spiral galaxies, subdivided 
by Hubble type and bar morphology.  The right axis scale shows corresponding
values of the stellar birthrate parameter $b$, which is the ratio of the
present SFR to that averaged over the past, as described in Section 5.1.}
\end{figure}

Part of the large dynamic range in absolute SFRs simply reflects the 
enormous range in galaxy masses, so it is more illuminating to examine
the range in relative SFRs, normalized per unit mass or luminosity.  This is 
illustrated in Figure 3, which shows 
the distribution of \halpha$+$[NII] equivalent widths (EWs) in a
sample of 227 nearby bright galaxies ($B_T<13$), subdivided by Hubble type.  
The data were taken from the photometric surveys of Kennicutt \& Kent
(1983) and Romanishin (1990).  The measurements include the \halpha\ and
the neighboring [NII] lines; corrections for [NII] contamination
are applied when determining the SFRs.  The EW 
is defined as the emission-line luminosity normalized to the adjacent
continuum flux, and hence it is a measure of the SFR per unit (red)
luminosity.  

Figure 3 shows a range of more than two orders of magnitude in the 
SFR per unit luminosity.  The EWs show a strong  
dependence on Hubble type, increasing from zero in E/S0 galaxies (within
the observational errors) to 20--150~\AA\ in late-type spiral and
irregular galaxies.  When expressed in terms of absolute SFRs, this
corresponds to range of 0--10 M$_\odot$~yr$^{-1}$ for an 
$L^*$ galaxy (roughly comparable in luminosity to the Milky Way).  The SFR 
measured in this way increases by approximately a factor of 20 between types 
Sa and Sc (Caldwell et al 1991, Kennicutt et al 1994).  SFRs derived from 
the UV continuum and broadband visible colors show comparable behavior 
(e.g. Larson \& Tinsley 1978, Donas et al 1987, 
Buat et al 1989, Deharveng et al 1994).  

High-resolution imaging of individual galaxies reveals
that the changes in the disk SFR along the Hubble
sequence are produced in roughly equal parts by an increase in the
total number of star forming regions per unit mass or area, and an
increase in the characteristic masses of individual regions (Kennicutt et al
1989a, Caldwell et al 1991, Bresolin \& Kennicutt 1997).  These trends
are seen both in the \halpha\ luminosities of the HII regions as well
as in the continuum luminosity functions of the embedded OB associations
(Bresolin \& Kennicutt 1997).  A typical OB star in an Sa galaxy forms in 
a cluster containing only a few massive stars, whereas an average 
massive star in a large Sc or Irr galaxy forms in a giant HII/OB
association containing hundreds or thousands of OB stars.  These differences
in clustering properties of the massive stars may strongly influence
the structure and dynamics of the interstellar medium (ISM) 
along the Hubble sequence (Norman \&
Ikeuchi 1989, Heiles 1990).

Although there is a strong trend in the {\it average} SFRs
with Hubble type, a dispersion of a factor of ten is present 
in SFRs among galaxies of the same type.  The scatter 
is much larger than would be expected from observational
errors or extinction effects, so most of it must 
reflect real variations in the SFR.  Several factors contribute to the SFR 
variations, including variations in gas content, nuclear emission,
interactions, and possibly short-term variations in the SFR within 
individual objects.  Although the absolute SFR varies considerably among
spirals (types Sa and later), some level of massive star formation is 
always observed in deep \halpha\ images 
(Caldwell et al 1991).  However many of the earliest disk galaxies (S0--S0/a)
show no detectable star formation at all.  Caldwell et al (1994) obtained
deep Fabry-Perot \halpha\ imaging of 8 S0--S0/a galaxies, and detected
HII regions in only 3 objects.  The total SFRs in the latter galaxies
are very low, $<$0.01~M$_\odot$~yr$^{-1}$, and the upper limits on the
other 4 galaxies rule out HII regions fainter than those of the Orion
nebula.  On the other hand, \halpha\ surveys of
HI-rich S0 galaxies by Pogge \& Eskridge (1987, 1993) reveal a higher
fraction of disk and/or circumnuclear star forming regions, emphasizing
the heterogeneous star formation properties of these galaxies.
Thronson et al (1989) reached similar conclusions based on an analysis
of IRAS observations of S0 galaxies.

The relative SFRs can also be parametrized in 
terms of the mean SFR per unit disk area.  This has the advantage of
avoiding any effect of bulge contamination on total luminosities 
(which biases the EW distributions). 
Analyses of the SFR surface density distributions have
been published by Deharveng et al (1994), based on UV continuum observations,
and by Ryder (1993), Ryder \& Dopita (1994), and Young et al (1996), 
based on \halpha\ observations.  The average SFR surface densities show
a similar increase with Hubble type, but the magnitude of the change is
noticeably weaker than is seen in SFRs per unit luminosity (e.g. Figure 3), 
and the dispersion among galaxies of the same type is larger (see below).
The stronger type dependence in the \halpha\
EWs (see Figure 3) is partly due to the effects of bulge contamination, 
which exaggerate the change in 
{\it disk} EWs by a factor of two between types Sa--Sc (Kennicutt et al 
1994), but the change in disk EWs with type is still nearly twice as large as 
the comparable trend in SFR per unit area (Young et al 1996).  The difference 
reflects the tendency for the late-type spirals to have somewhat more extended
(i.e. lower surface brightness) star forming disks than the early-type
spirals, at least in these samples.  This comparison demonstrates 
the danger in applying the term SFR too loosely when characterizing
the systematic behavior of star formation across the Hubble sequence, 
because the quantitative trends are dependent on the manner in which the 
SFR is defined.  Generally speaking, a parameter that scales with the 
SFR per unit mass (e.g. the \halpha\ equivalent width) is most relevant
to interpreting the evolutionary properties of disks, whereas
the SFR per unit area is more relevant to parametrizing the dependence
of the SFR on gas density in disks.

\begin{figure}[!ht]
  \begin{center}
    \leavevmode
  \centerline{\epsfig{file=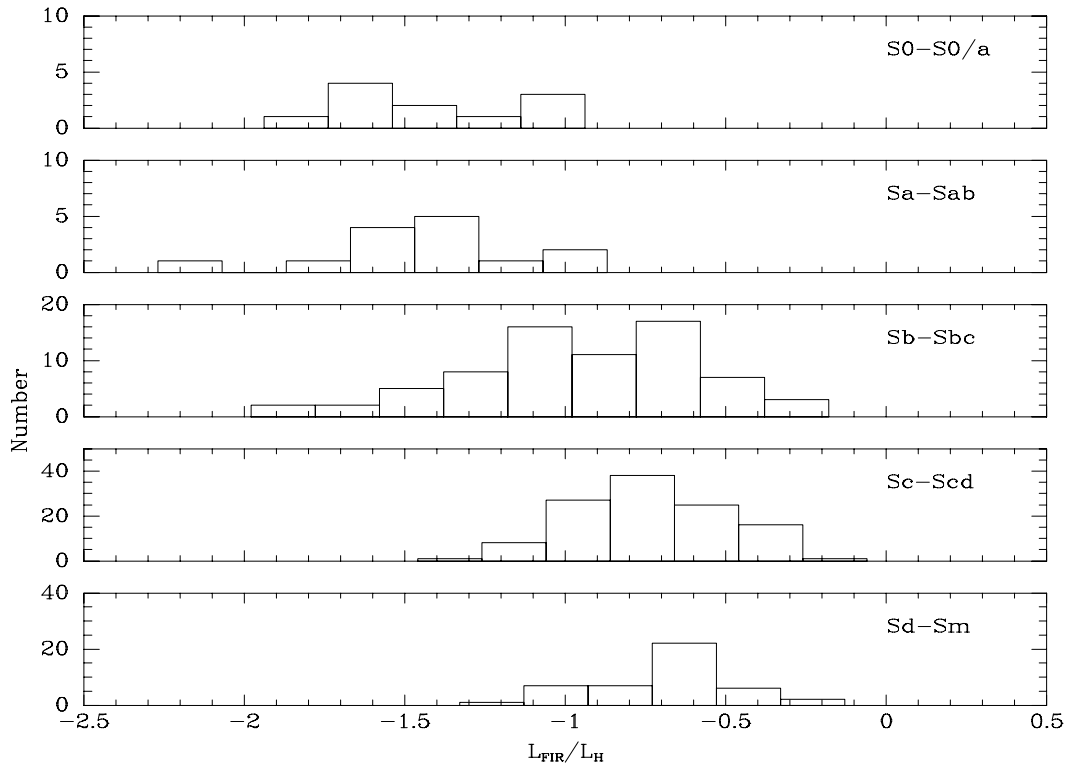,width=16cm}}
  \end{center}
   \caption{\em  Distributions of 40- to 120-$\mu$m infrared luminosity for 
nearby galaxies, normalized to near-infrared $H$ luminosity, as a function
of Hubble type.  Adapted from Devereux \& Hameed (1997), with elliptical
and irregular galaxies excluded.} 
  \end{figure}

Similar comparisons can be made for the FIR properties of disk galaxies,
and these show considerably weaker trends with Hubble type (Devereux
\& Young 1991, Tomita et al 1996, Devereux \& Hameed 1997).  This is 
illustrated in Figure 4, which shows the distributions of $L_{FIR}$/$L_H$
from a sample of nearby galaxies studied by Devereux \& Hameed (1997).
Since the near-IR $H$-band luminosity is a good indicator of the total
stellar mass, the L$_{FIR}$/L$_H$ ratio provides an approximate measure of the 
FIR emission normalized to the mass of the parent galaxy.
Figure 4 shows the expected trend toward stronger FIR emission with
later Hubble type, but the trend is considerably weaker, in the sense that
early-type galaxies show much higher FIR luminosities than would be 
expected given their UV-visible spectra. 
Comparisons of L$_{FIR}$/L$_B$ distributions show 
almost no dependence on Hubble type at all (Isobe \& Feigelson 1992, 
Tomita et al 1996, Devereux \& Hameed 1997), but this is 
misleading because the $B$-band luminosity itself correlates with the SFR
(see Figure 2).

The inconsistencies between the FIR and UV--visible properties of
spiral galaxies appear to be due to a combination of effects (as
mentioned above in Section 2.5).  In at 
least some early-type spirals, the strong FIR emission is produced by
luminous, dusty star forming regions, usually concentrated in the
central regions of barred spiral galaxies (Devereux 1987, Devereux
\& Hameed 1997).  This exposes an 
important bias in the visible and UV-based studies of SFRs in galaxies,
in that they often do not take into account the substantial star formation
in the dusty nuclear regions, which can dominate 
the global SFR in an early-type galaxy.  Devereux \& Hameed
emphasize the importance of observing a sufficiently large and diverse
sample of early-type galaxies, in order to fully characterize the range
of star formation properties.  However it is also 
likely that much of the excess FIR emission in early-type spirals 
is unrelated to star formation, reflecting instead the effects of 
dust heating from evolved stellar populations (Section 2.5).  
Radiative transfer modelling by Walterbos \& Greenawalt (1996)
demonstrates that this effect can readily account for the trends
seen in Figure 4.  

The interpretation in the remainder of this review is based on the 
SFR trends revealed by the \halpha, UV continuum, broadband colors,
and integrated spectra, which are consistent with a common evolutionary
picture of the Hubble sequence.  However it is important to bear
in mind that this picture applies only to the extended, extranuclear
star formation in spiral and irregular disks.  The circumnuclear star
formation follows quite different patterns, as discussed in Section 4.2.

\subsection{{\it Dependence of SFRs on Gas Content}}

The strong trends in disk SFRs that characterize the Hubble sequence
presumably arise from more fundamental relationships between the global SFR
and other physical properties of galaxies, such as their 
gas contents or dynamical structure.  The  
physical regulation of the SFR is a mature subject in its own right,
and a full discussion is beyond the scope of this review.  However
it is very instructive to examine the global relationships between 
the disk-averaged SFRs and gas densities of galaxies, because they 
reveal important insights into the physical nature of the star 
formation sequence, and they serve to quantify the range of physical
conditions and evolutionary properties of disks.

Comparisons of the large-scale SFRs and gas contents of galaxies
have been carried out by several authors, most recently Buat et al
(1989), Kennicutt (1989), Buat (1992), Boselli (1994), Deharveng
et al (1994), Boselli et al (1995) and Kennicutt (1998).  Figure 5 shows 
the relationship between the disk-averaged SFR surface density 
$\Sigma_{SFR}$ and average total
(atomic plus molecular) gas density $\Sigma_{gas}$, for a sample of 61 normal
spiral galaxies with \halpha, HI, and CO observations (Kennicutt 1998).  
The SFRs were derived from extinction-corrected \halpha\ fluxes, 
using the SFR calibration in Equation 2.  The surface densities were
averaged within the corrected optical radius $R_0$, as taken from the 
{\it Second Reference Catalog of Bright Galaxies} (de Vaucouleurs et al 1976).

\begin{figure}[!ht]
  \begin{center}
    \leavevmode
  \centerline{\epsfig{file=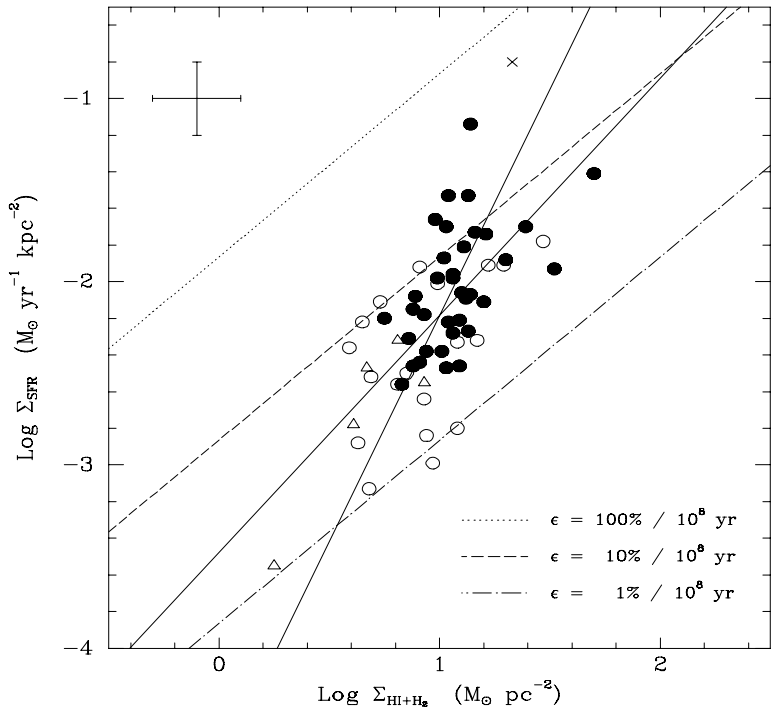,width=16cm}}
  \end{center}
  \caption{\em Correlation between disk-averaged SFR per unit area and 
average gas surface density, for 61 normal disk galaxies.  Symbols are coded
by Hubble type:  Sa--Sab (open triangles); Sb--Sbc (open circles);
Sc--Sd (solid points); Irr (cross).  The dashed and dotted lines show 
lines of constant global star formation efficiency.  The error bars
indicate the typical uncertainties for a given galaxy, including 
systematic errors.}
   \end{figure}

Figure 5 shows that disks possess large ranges in both the mean 
gas density (factor of 20--30) and mean SFR surface density (factor of 100).  
The data points are coded by galaxy type, and they show that both the gas 
and SFR densities are correlated with Hubble type on average, but with 
large variations among galaxies of a given type.  In addition, there
is an underlying correlation between SFR and gas density that is largely 
independent of galaxy type.  This shows that 
much of the scatter in SFRs among galaxies of the same type can be attributed
to an underlying dispersion in gas contents.  The data can be fitted
to a Schmidt (1959) law of the form\ \ $\Sigma_{SFR} = A~\Sigma{_{gas}^N}$.
The best fitting slope $N$ ranges from 1.4 for a conventional least squares
fit (minimizing errors in SFRs only) to $N$=2.4 for a bivariate regression,
as shown by the solid lines in Figure 5.  Values of $N$ in the
range 0.9--1.7 have been derived by previous workers, based on SFRs
derived from \halpha, UV, and FIR data (Buat et al 1989, Kennicutt 1989,
Buat 1992, Deharveng et al 1994).  
The scatter in SFRs at a given gas density is large, and most of this 
dispersion is probably introduced by averaging the SFRs and gas densities
over a large dynamic range of local densities within the individual
disks (Kennicutt 1989, 1998).  

Figure 5 also contains information on the typical global efficiencies
of star formation and gas consumption time scales in disks.  The dashed
and dotted lines indicate constant, disk-averaged efficiencies of 1\%, 
10\%, and 100\% per $10^8$ years.  The average value for these galaxies
is 4.8\%, meaning that the average disk converts
4.8\%\ of its gas (within the radius of the optical disk) every $10^8$ years.
Since the typical gas mass fraction in these disk is about 20\%,
this implies that stellar mass of the disk grows by about 1\%\ per
$10^8$ years, i.e. the time scale for building the disk (at the present
rate) is comparable to the Hubble time.  The
efficiencies can also be expressed in terms of the average gas 
depletion timescale, which for this sample is 2.1~Gyr.  Recycling of 
interstellar gas from stars extends the actual time scale for gas depletion by
factors of 2--3 (Ostriker \& Thuan 1975, Kennicutt et al 1994).

\subsection{{\it Other Global Influences on the SFR}}

What other global properties of a galaxy influence its SFR?
One might plausibly expect the mass, bar structure, spiral arm structure,
or environment to be important, and empirical information on all of
these are available.

3.3.1~~LUMINOSITY AND MASS~~~Gavazzi \& Scodeggio (1996) and Gavazzi et al 
(1996) have compiled UV, visible, and near-IR photometry for over 900 
nearby galaxies, and they found  
an anti-correlation between the SFR per unit mass and the galaxy 
luminosity, as indicated by broadband colors and \halpha\ EWs.  At least part 
of this trend seems to reflect the same dependence of SFR on Hubble type 
discussed above, but a mass dependence is also observed among galaxies
of the same Hubble type.  
It is interesting that there is considerable
overlap between the color-luminosity relations of different spiral types,
which suggests that part of the trends that are attributed to morphological
type may be more fundamentally related to total mass.  A strong correlation
between $B$--$H$ color and galaxy luminosity or linewidth has been discussed
previously by Tully et al (1982) and Wyse (1983).  The trends 
seem to be especially strong for redder colors, which are more closely
tied to the star formation history and mean metallicity than the current 
SFR.  More data are needed to fully disentangle the effects of galaxy
type and mass, both for the SFR and the star formation history.

3.3.2~~BARS~~~Stellar bars can strongly perturb the gas flows in disks,
and trigger nuclear star formation (see next section), but they do not appear
to significantly affect the total disk SFRs.  
Figure 3 plots the \halpha\ EW distributions
separately for normal (SA and SAB) and barred (SB) spirals,
as classified in the {\it Second Reference Catalog of Bright Galaxies}.  There
is no significant difference in the EW distributions (except possibly
for the Sa/SBa galaxies), which suggests
that the global effect of a bar on the {\it disk} SFR is unimportant.
Ryder \& Dopita (1994) reached the same conclusion based on \halpha\
observations of 24 southern galaxies.  

Tomita et al (1996) have carried out a similar comparison of FIR emission,
based on IRAS data and broadband photometry for 139
normal spirals and 260 barred Sa--Sc galaxies.  They compared the
distributions of $L_FIR$/$L_B$ ratios for Sa/SBa, Sb/SBb, and Sc/SBc
galaxies, and concluded that there is no significant correlation with
bar structure, consistent with the \halpha\ results.  There is
evidence for a slight excess in FIR emission in SBa galaxies, which could
reflect bar-triggered circumnuclear star formation in some of the galaxies, 
though the statistical significance of 
the result is marginal (Tomita et al 1996). 

Recent work by Martinet \& Friedli (1997) suggests that influence of 
bars on the global SFR may not be as simple as suggested above.  
They analyzed \halpha\
and FIR-based SFRs for a sample of 32 late-type barred galaxies, and
found a correlation between SFR and the strength and length of the bar.
This suggests that large samples are needed to study the effects of
bars on the large-scale SFR, and that the structural properties of
the bars themselves need to be incorporated in the analysis.

3.3.3~~SPIRAL ARM STRUCTURE~~~Similar tests have been carried out to 
explore whether a strong 
grand-design spiral structure enhances the global SFR.  Elmegreen \&
Elmegreen (1986) compared UV and visible broadband colors and \halpha\
EWs for galaxies they classified as grand-design (strong two-armed
spiral patterns) and flocculent (ill-defined, patchy spiral arms),
and they found no significant difference in SFRs.  McCall \& Schmidt (1986)
compared the supernova rates in grand-design and flocculent spirals, and
drew similar conclusions.  Grand-design spiral galaxies
show strong local enhancements of star formation in their spiral arms
(e.g. Cepa \& Beckman 1990, Knapen et al 1992), 
so the absence of a corresponding excess
in their total SFRs suggests that the primary effect of the spiral
density wave is to concentrate star formation in the arms, but not
to increase the global efficiency of star formation.

3.3.4~~GALAXY-GALAXY INTERACTIONS~~~Given the modest effects of internal 
disk structure on global SFRs,
it is perhaps somewhat surprising that external environmental 
influences can have much stronger effects on the SFR.  The most
important influences by far are tidal interactions.  Numerous
studies of the global \halpha\ and FIR emission of interacting
and merging galaxies have shown strong excess star formation
(e.g. Bushouse 1987, Kennicutt et al 1987, Bushouse et al 1988, 
Telesco et al 1988, Xu \& Sulentic 1991, Liu \& Kennicutt 1995).
The degree of the SFR enhancement is highly variable, ranging
from zero in gas-poor galaxies to on the order of 10--100 times in extreme
case.  The average enhancement in SFR over large samples is a factor
of 2--3.  Much larger enhancements are often seen
in the circumnuclear regions of strongly interacting and merging
systems (see next section).  This subject is reviewed in depth in 
Kennicutt et al (1998).  

3.3.5~~CLUSTER ENVIRONMENT~~~There is evidence that cluster environment
systematically alters the star formation properties of galaxies, independently
of the well-known density-morphology relation (Dressler 1984).
Many spiral galaxies located in rich clusters
exhibit significant atomic gas deficiencies (Haynes et al 1984, 
Warmels 1988, Cayatte et al 1994), which presumably are the result of 
ram pressure
stripping from the intercluster medium, combined with tidal stripping
from interactions with other galaxies and the cluster potential.
In extreme cases one would expect the gas removal to affect the 
SFRs as well.  Kennicutt (1983b) compared \halpha\ EWs of 26 late-type
spirals in the Virgo cluster core with the field sample of Kennicutt
\& Kent (1983) and found evidence for a 50\%\ lower SFR in Virgo,
comparable to the level of HI deficiency.  The UV observations of
the cluster Abell 1367 by Donas et al (1990) also show evidence for
lower SFRs.  However subsequent studies
of the Coma, Cancer, and A1367 clusters by Kennicutt et al (1984) 
and Gavazzi et al (1991) showed no reduction in the average SFRs, and if
anything a higher number of strong emission-line galaxies.

A comprehensive \halpha\ survey of 8 Abell clusters by Moss \& Whittle 
(1993) suggests that the effects of cluster environoment on global
star formation activity are quite complex.  They found a 37--46\%\
lower \halpha\ detection rate among Sb, Sc, and irregular galaxies
in the clusters, but a 50\%\ higher detection rate among Sa--Sab galaxies.
They argue that these results arise from a combination of
competing effects, including reduced star formation from gas stripping
as well as enhanced star formation triggered by tidal interactions.
Ram-pressure induced star formation may also be taking place in a 
few objects (Gavazzi \& Jaffe 1985).  

\section{CIRCUMNUCLEAR STAR FORMATION \\ 
AND STARBURSTS} 

It has been known from the early
photographic work of Morgan (1958) and S\'ersic \& Pastoriza (1967) that the 
circumnuclear regions of many spiral galaxies harbor 
luminous star forming regions, with properties that are largely decoupled 
from those of the more extended star forming disks.  
Subsequent spectroscopic surveys revealed numerous
examples of bright emission-line nuclei with spectra resembling
those of HII regions (e.g. Heckman et al 1980, Stauffer 1982, 
Balzano 1983, Keel 1983).  The most luminous of these were dubbed
``starbursts" by Weedman et al (1981).  The opening of the 
mid-IR and FIR regions  
fully revealed the distinctive nature of the nuclear star formation
(e.g. Rieke \& Low 1972, Harper \& Low 1973, Rieke \& Lebofsky 1978,
Telesco \& Harper 1980).  The IRAS survey 
led to the discovery of large numbers of ultraluminous star forming galaxies
(Soifer et al 1987).  This subject has grown into a major 
subfield of its own, which has been thoroughly reviewed elsewhere in this
series (Soifer et al 1987, Telesco 1988, Sanders \& Mirabel 1996).
The discussion here focusses on the range of star formation properties of the
nuclear regions, and the patterns in these properties along the 
Hubble sequence.

\subsection{{\it SFRs and Physical Properties}}

Comprehensive surveys of the star formation properties of galactic
nuclei have been carried out using emission-line spectroscopy in
the visible (Stauffer 1982, Keel 1983, Kennicutt et al 1989b, Ho et al 1997a, b)
and mid-IR photometry (Rieke \& Lebofsky 1978, Scoville et al 
1983, Devereux et al
1987, Devereux 1987, Giuricin et al 1994).  Nuclear emission spectra with
HII region-like line ratios are found in 42\%\ of bright spirals
($B_T < 12.5$), with the fraction increasing from 8\%\ in S0 galaxies
(and virtually zero in elliptical galaxies) 
to 80\%\ in Sc--Im galaxies (Ho et al 1997a).  These fractions are
lower limits, especially in early-type spirals, because the star formation
often is masked by a LINER or Seyfert nucleus.
Similar detection fractions are found in 10$\mu$m surveys of 
optically-selected spiral galaxies, but with a stronger weighting toward
early Hubble types.  The  
nuclear SFRs implied by the \halpha\ and IR fluxes span a large range,
from a lower detection limit 
of $\sim$10$^{-4}$~M$_\odot$~yr$^{-1}$ to well over 100~\sfr\ in
the most luminous IR galaxies.  

The physical character of the nuclear star forming regions changes
dramatically over this spectrum of SFRs.  The nuclear SFRs in most 
galaxies are quite modest, averaging $\sim$0.1~\sfr\ (median 0.02~\sfr) 
in the \halpha\ sample of Ho et al (1997a), and $\sim$0.2~\sfr\ in
the (optically selected) 10$\mu$m samples of  
Scoville et al (1983) and Devereux et al (1987).  Given the 
different selection criteria and completeness levels in these surveys, 
the SFRs are reasonably consistent with each other, and this suggests that 
the nuclear star formation at the low end of the SFR spectrum typically
occurs in moderately obscured regions ($A_{H\alpha}\sim$0--3 mag) 
that are not physically dissimilar from normal disk HII regions 
(Kennicutt et al 1989b, Ho et al 1997a).  

However the IR observations also reveal a population of more luminous 
regions, with $L_{FIR} \sim 10^{10}$--$10^{13}$~$L_\odot$, and corresponding
SFRs of order 1--1000~\sfr\ (Rieke \& Low 1972, Scoville et al 1983,
Joseph \& Wright 1985, Devereux 1987).  Such high SFRs are not seen
in optically-selected samples, mainly because the luminous starbursts
are uniquely associated with dense molecular gas disks (Young \& Scoville
1991 and references therein), and for normal gas-to-dust ratios, one expects
visible extinctions of several magnitudes or higher.  The remainder of this 
section will 
focus on these luminous nuclear starbursts, because they represent a
star formation regime that is distinct from the more extended star formation 
in disks, and because these bursts often dominate the total
SFRs in their parent galaxies.

\begin{figure}[!ht]
  \begin{center}
    \leavevmode
  \centerline{\epsfig{file=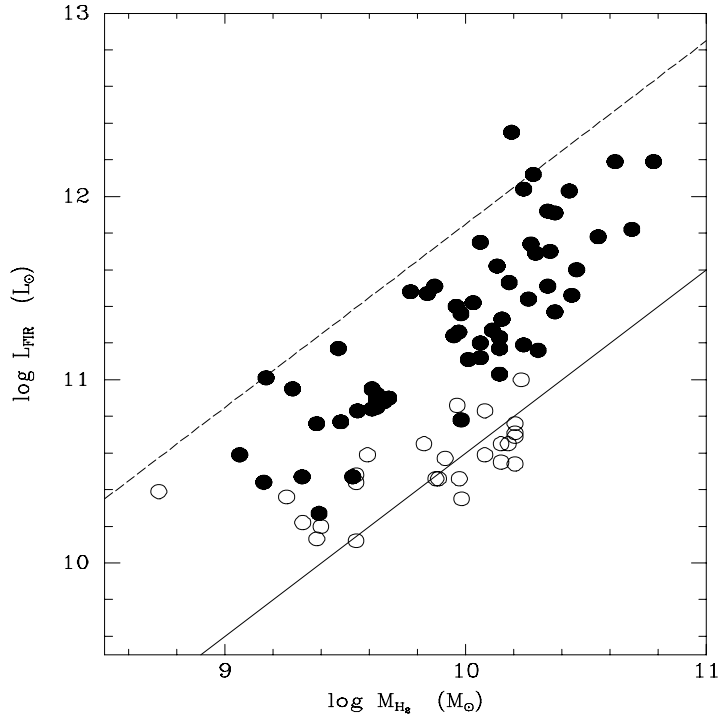,width=16cm}}
  \end{center}
  \caption{\em Relationship between integrated far-infrared (FIR) 
luminosity and 
molecular gas mass for bright IR galaxies, from Tinney et al (1990;
open circles) and a more luminous sample from Sanders et al (1991; solid
points).  The solid line shows the typical $L/M$ ratio for galaxies similar to
the Milky Way.  The dashed line shows the approximate limiting luminosity 
for a galaxy forming stars with 100\%\ efficiency on a dynamical timescale, 
as described in the text.}  
\end{figure}

The IRAS all-sky survey provided the first comprehensive picture
of this upper extreme in the SFR spectrum.  
Figure 6 shows a comparison of the total 8--1000~$\mu$m luminosities  
(as derived from IRAS) and total molecular gas masses for 87 
IR-luminous galaxies, taken from the surveys of Tinney et al (1990) and 
Sanders et al (1991).  Tinney et al's sample (open circles) includes
many luminous but otherwise normal star forming galaxies, while 
Sanders et al's brighter sample (solid points) mainly comprises starburst 
galaxies and a few AGNs.  Strictly speaking these measurements cannot
be applied to infer the nuclear SFRs of the galaxies, because they are
low-resolution measurements and the samples are heterogeneous.  However
circumnuclear star formation sufficiently dominates the properties of 
the luminous infrared galaxies (e.g. Veilleux et al 1995, Lutz et al 1996)
that Figure 6 (solid points) provides a representative indication of
the range of SFRs in these IRAS-selected samples.

The most distinctive feature in Figure 6 is the range of infrared luminosities.
The lower range overlaps with the luminosity function of normal galaxies
(the lower limit of $10^{10}~L_\odot$ is the sample definition 
cutoff), but the population of infrared galaxies extends upward to  
$>$10$^{12.5}~L_\odot$.  This would imply SFRs of up to 
500 \sfr\ (Equation 4), if starbursts are primarily 
responsible for the dust heating, about 20 times larger than the highest
SFRs observed in normal galaxies.  Figure 6 also shows that the luminous 
IR galaxies are associated with unusally high molecular gas masses, 
which partly accounts for the high SFRs.  However the typical SFR per unit 
gas mass is much higher than in normal disks; the solid
line in Figure 6 shows the typical $L/M$ ratio for normal galaxies, 
and the efficiencies in the IR galaxies are higher
by factors of 2--30 (Young et al 1986, Solomon \& Sage 1988, 
Sanders et al 1991).  The H$_2$ masses used here have been
derived using a standard Galactic H$_2$/CO conversion ratio, and
if the actual conversion factor in the IRAS galaxies is lower,
as is suggested by several lines of evidence, the contrast in 
star formation efficiencies would be even larger (e.g. Downes et al 1993, 
Aalto et al 1994, Solomon et al 1997).

High-resolution IR photometry and imaging of the luminous infrared 
galaxies reveals that the bulk of the luminosity originates
in compact circumnuclear regions (e.g. Wright et al 1988, Carico et al 1990, 
Telesco et al 1993, Smith \& Harvey
1996, and references therein).  Likewise, CO interferometric observations 
show that a large fraction of the molecular gas is concentrated in central 
disks, with typical radii on the order of 0.1--1 kpc, and implied surface
densities on the order of $10^2$--$10^5~M_\odot$~pc$^{-2}$ (Young \& Scoville
1991, Scoville et al 1994, Sanders \& Mirabel 1996).  Less massive disks
but with similar gas and SFR surface densities are associated with 
the infrared-bright nuclei of spiral galaxies 
(e.g. Young \& Scoville 1991, Telesco et al 1993, Scoville et al 
1994, Smith \& Harvey 1996, Rubin et al 1997).  The full spectrum of
nuclear starburst regions will be considered in the remainder of this section.

\begin{table}
\caption{Star Formation in Disks and Nuclei of Galaxies}

\bigskip
\begin{tabular}{lcc}\hline\hline \\

\bigskip

Property& Spiral Disks & Circumnuclear Regions \\

\hline 
 & & \\
Radius  &  $1 - 30$ kpc & $0.2 - 2$ kpc \\
SFR &      $0 - 20$ M$_\odot$~yr$^{-1}$ & $0 - 1000$ M$_\odot$~yr$^{-1}$ \\
Bolometric Luminosity   & $10^6 - 10^{11}$ L$_\odot$ & $10^6 - 10^{13}$ L$_\odot$ \\
Gas Mass   & $10^8 - 10^{11}$ M$_\odot$ & $10^6 - 10^{11}$ M$_\odot$ \\
Star Formation Timescale  & $1 - 50$ Gyr & $0.1 - 1$ Gyr \\
Gas Density  & $1 - 100$ M$_\odot$~pc$^{-2}$ & $10^2 - 10^5$ M$_\odot$~pc$^{-2}$ \\
Optical Depth (0.5 $\mu$m) & $0 - 2$ & $1 - 1000$ \\
SFR Density  & $0 - 0.1$ M$_\odot$~yr$^{-1}$~kpc$^{-2}$ & $1 - 1000$ M$_\odot$~yr$^{-1}$~kpc$^{-2}$ \\
Dominant Mode & steady state & steady state $+$ burst \\
 & & \\
\hline
\\

Type Dependence? & strong & weak/none \\
Bar Dependence? & weak/none & strong \\
Spiral Structure Dependence?  & weak/none & weak/none \\
Interactions Dependence? & moderate & strong \\
Cluster Dependence? & moderate/weak & ? \\
Redshift Dependence? & strong & ? \\

\\ 

\hline
\end{tabular}
\end{table}

The physical conditions in the circumnuclear star forming disks
are distinct in many respects from the more extended star forming disks
of spiral galaxies, as is summarized in Table 1.  The circumnuclear 
star formation is especially distinctive in terms of the
absolute range in SFRs, the the much higher spatial
concentrations of gas and stars, its burstlike nature (in luminous 
systems), and its systematic variation with galaxy type.

The different range of physical conditions in the nuclear starbursts
is clearly seen in Figure 7, which plots the average SFR surface densities
and mean molecular surface densities for the circumnuclear disks of 
36 IR-selected starbursts (Kennicutt 1998).  The comparison
is identical to the SFR--density plot for spiral disks in Figure 5,
except that in this case the SFRs are derived from FIR luminosities 
(equation [4]), and only molecular gas densities  
are plotted.  HI observations show that the atomic gas fractions in these 
regions are of the order of only a few percent, and can be safely 
neglected (Sanders \& Mirabel 1996).  The SFRs and densities have
been averaged over the radius of the circumnuclear disk, as measured
from high-resolution CO or IR maps (see Kennicutt 1998).

\begin{figure}[!ht]
  \begin{center}
    \leavevmode
  \centerline{\epsfig{file=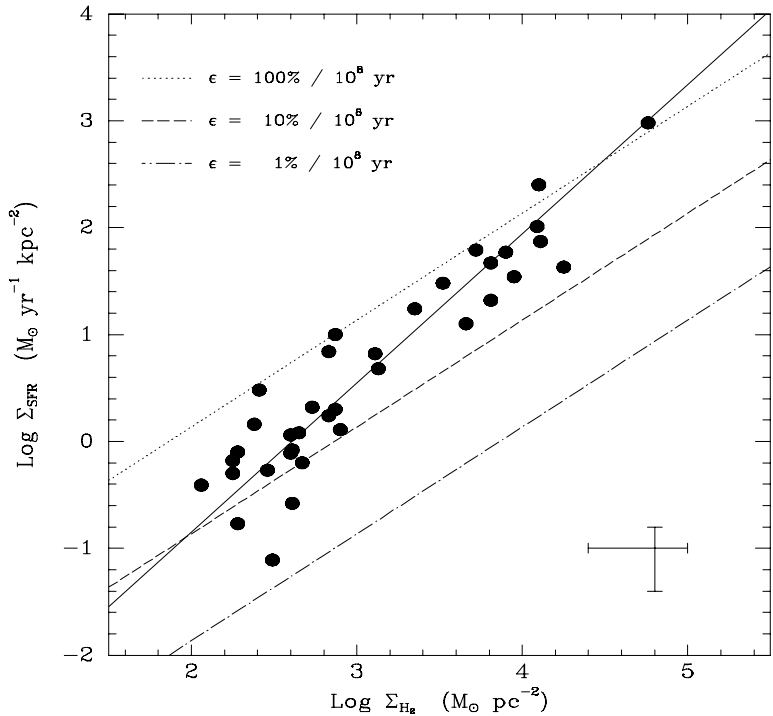,width=16cm}}
  \end{center}
  \caption{\em Correlation between disk-averaged SFR per unit area and 
average gas surface density, for 36 infrared-selected circumnuclear starbursts. 
See Figure 5 for a similar comparison for normal spiral disks.  The dashed 
and dotted lines show lines of constant star formation conversion efficiency,
with the same notation as in Figure 5.  The error bars indicate the typical
uncertainties for a given galaxy, including systematic errors.}
\end{figure}

Figure 7 shows that the surface densities of gas and star formation
in the nuclear starbursts are 1--4 orders of magnitude higher than in 
spiral disks overall.  Densities of this order can be found in large
molecular cloud complexes within spiral disks, of course,
but the physical conditions in many
of the nuclear starbursts are extraordinary even by those standards.
For example, the typical mean densities of the largest molecular 
cloud complexes in M31, M33, and M51 are in the range 40--500
$M_\odot$~pc$^{-2}$, which corresponds to the lower range of 
densities in Figure 7 (Kennicutt 1998).  Likewise the SFR surface
densities in the 30 Doradus giant HII region, the most luminous
complex in the Local Group, reaches 100~$M_\odot$~yr$^{-1}$~kpc$^{-2}$
only in the central 10 pc core cluster.  The corresponding densities
in many of the starbursts exceed these values, over
regions as large as a kiloparsec in radius.

The starbursts follow a relatively well-defined Schmidt law, with
index $N$$\sim$1.4.  The nature of the star formation law will be
discussed further in Section 5, where we examine the SFR {\it vs} gas 
density relation for all of the data taken together.
Figure 7 also shows that the characteristic star formation efficiencies
and timescales are quite different in the starbursts.  The mean conversion 
efficiency is 30\%\ per $10^8$ years, 6 times larger than in the spiral disks.
Likewise, the gas consumption timescale is 6 times shorter, about 0.3 Gyr
on average.  This is hardly surprising--- these objects are 
starbursts by definition--- but Figure 7 serves to 
quantify the characteristic timescales for the starbursts.  

As pointed out by Heckman (1994) and Lehnert \& Heckman (1996),
the luminous IR galaxies lie close to the limiting luminosity allowed
by stellar energy generation, for a system which converts all of its
gas to stars over a dynamical timescale.  For a galaxy with dimensions
comparable to the Milky Way, the minimum timescale for feeding the 
central starburst is $\sim$10$^8$ years; this is also consistent with
the minimum gas consumption timescale in Figure 7.  At the limit of 100\%\ 
star formation efficiency over this timescale, the corresponding SFR is 
trivially:
\begin{equation}
{{\rm SFR}_{max}} = { {100~M_\odot~{\rm yr^{-1}}}~{({{M_{gas}} \over {10^{10}~M_\odot}})}~{({{10^8~years} \over {\tau_{dyn}}})} }. 
\end{equation}

The corresponding maximum bolometric luminosity can be estimated using
Equation 4, or by calculating the maximum nuclear energy release
possible from stars over $10^8$ years.  The latter is  
$\sim$0.01~$\epsilon \dot{m} c^2$, where $\dot{m}$ 
in this case is the SFR, and $\epsilon$ is the fraction of the total stellar
mass that is burned in $10^8$ yr.  A reasonable value of $\epsilon$ for a 
Salpeter IMF is about 0.05; it could be as high as 0.2 if the starburst
IMF is depleted in low-mass stars (e.g. Rieke et al 1993). 
Combining these terms and assuming further
that all of the bolometric luminosity is reradiated by the dust yields:
\begin{equation}
{L_{max}} \sim {{7 \times 10^{11}~L_\odot}~{({M_{gas} \over {10^{10}~M_\odot}})}~
{({\epsilon \over 0.05})}}.
\end{equation}

Using Equation 4 to convert the SFR to FIR luminosity gives
nearly the same coefficient ($6 \times 10^{11}$).
This limiting $L/M$ relation is shown by the dashed line
in Figure 6, and it lies very close to the actual upper envelope of
the luminous IR galaxies.  Given the number of assumptions that
went into Equation 6, this agreement may be partly fortuitous;  
other physical processes, such as optical depth
effects in the cloud, may also be important in defining the upper luminosity
limits (e.g. Downes et al 1993).  However the main intent of this exercise
is to illustrate that many of the most extreme circumnuclear starbursts
lie near the physical limit for maximum SFRs in galaxies.  
Heckman (1994) extended this argument and derived the
maximum SFR for a purely self-gravitating protogalaxy, and he showed 
that the most luminous infrared galaxies lie close to this limit
as well.  Note that none of these limits apply to AGN-powered galaxies,
because the mass consumption requirements for a given mass are
1--2 orders of magnitude lower.

Taken together, these results reveal the extraordinary 
character of the most luminous IR starburst galaxies (Heckman 1994,
Scoville et al 1994, Sanders \& Mirabel 1996).  They 
represent systems in which a mass of gas comparable to the entire 
ISM of a galaxy has been driven into a region of order 1 kpc
in size, and this entire ISM is being formed into stars, with almost
100\%\ efficiency, over a timescale of order $10^8$ years.
Such a catastrophic transfer of mass can only
take place in a violent interaction or merger, or perhaps during
the initial collapse phase of protogalaxies.  

\subsection{{\it Dependence on Type and Environment}}

The star formation that takes place in the circumnuclear regions of
galaxies also follows quite different patterns along the Hubble sequence,
relative to the more extended star formation in disks.
These distinctions are especially important in early-type galaxies,
where the nuclear regions often dominate the global star formation
in their parent galaxies.

4.2.1~~HUBBLE TYPE~~~In contrast to the extended star formation
in disks, which varies dramatically along the Hubble sequence, 
circumnuclear star formation is largely decoupled with Hubble type.
Stauffer (1982), Keel (1983), and Ho et al (1997a) have investigated the 
dependence of nuclear \halpha\ emission in star forming nuclei as a function 
of galaxy type.  The detection frequency of HII region nuclei is
a strong monotonic function of type, increasing from 0\%\ in 
elliptical galaxies, to 8\%\ in SO, 22\%\ in Sa, 51\%\ in Sb, and
80\%\ in Sc--Im galaxies (Ho et al 1997a), though these fractions may be
influenced somewhat by AGN contamination.  
Among the galaxies with nuclear star formation, the \halpha\ 
luminosities show the opposite trend; the average extinction-corrected
luminosity of HII region nuclei in S0--Sbc galaxies is 9 times higher
than in Sc galaxies (Ho et al 1997a).  Thus the bulk of the total
nuclear star formation in galaxies is weighted toward the earlier
Hubble types, even though the frequency of occurence is higher in
the late types.

Similar trends are observed in 10~$\mu$m surveys of nearby galaxies
(Rieke \& Lebofsky 1978, Scoville et al 1983, Devereux et al 1987,
Devereux 1987, Giuricin et al 1994).  Interpreting the trends in 
nuclear 10$\mu$m luminosities by themselves is less straightforward,
because the dust can be heated by active nuclei as well as by star
formation, but one can reduce this problem by excluding known AGNs
from the statistics.  Devereux et al (1987) analyzed the properties
of an optically selected sample of 191 bright spirals, chosen
to lie within or near the distance of the Virgo cluster, and found
that the average nuclear 10$\mu$m flux was virtually independent of
type and if anything, decreased by 25--30\%\ from types Sa--Sbc
to Sc--Scd.  An analysis of a larger sample by Giuricin et al (1994)
shows that among galaxies with HII region nuclei (as classified from
optical spectra), Sa--Sb nuclei are 1.7 times more luminous
at 10$\mu$m than Sc galaxies.  By contrast  
the disk SFRs are typically 5--10 times lower in the early-type spirals, 
so the fractional contribution of the nuclei to the total SFR
increases dramatically in the early-type spirals.  The nuclear SFRs in some
early-type galaxies are comparable to the {\it integrated}
SFRs of late-type spirals (e.g. Devereux 1987, Devereux \& Hameed 1997).
Thus while luminous nuclear starbursts may occur in across the 
entire range of spiral host types (e.g. Rieke \& Lebofsky 1978, Devereux 1987),
the relative effect is much stronger for the early-type galaxies; most
of the star formation in these galaxies occurs in the circumnuclear regions.
Clearly the physical mechanisms that trigger these nuclear outbursts
are largely decoupled from the global gas contents and SFRs of their
parent galaxies.

4.2.2~~BAR STRUCTURE~~~These same surveys show that 
nuclear star formation is strongly correlated with the presence of
strong stellar bars in the parent galaxy.  
The first clear evidence came from the photographic work 
of S\'ersic \& Pastoriza (1967), who showed that 24\%\ of nearby 
SB and SAB galaxies possessed detectable circumnuclear ``hotspot"
regions, now known to be bright HII regions and stellar associations.
In contrast none of the non-barred galaxies studied showed evidence
for hotspots.  This work was followed up by Phillips (1993) who
showed that hotspots are found preferentially in early-type barred
galaxies, a tendency noted already by S\'ersic \& Pastoriza.  

The effects of bars on the \halpha\ emission from HII region nuclei
have been analyzed by Ho et al (1997b).  They found that the incidence of 
nuclear star formation is higher among the barred galaxies, but the difference
is marginally significant, and no excess is seen among early-type
barred galaxies.  However the distributions of \halpha\ luminosities
are markedly different, with the barred galaxies showing an extended
tail of bright nuclei that is absent in samples of non-barred galaxies.
This tail extends over a range in \halpha\ luminosities of 
3--100 $\times 10^{40}$ ergs~s$^{-1}$, which corresponds to
SFRs in the range 0.2--8~\sfr.  This tail is especially strong in
the early-type barred galaxies (SB0/a--SBbc), where $\sim$30\%\
of the star forming nuclei have luminosities in this range.

Bars appear to play an especially strong role in triggering the 
strong IR-luminous starbursts that are found in early-type spiral
galaxies.  Hawarden et al (1986) and Dressel (1988) found strong 
excess FIR emission in early-type barred spirals, based on IRAS
observations, and hypothesized that
this emission was associated with circumnuclear star forming regions.
This interpretation was directly confirmed by Devereux (1987), who
detected strong nuclear 10$\mu$m emission in 40\%\ of the early-type
barred spirals in his sample.  Similar excesses are not seen in
samples of late-type barred galaxies.  These results have been confirmed
in more extensive later studies by Giuricin et al (1994) and 
Huang et al (1996).  Although early-type barred galaxies frequently
harbor a bright nuclear starburst, bars are not a necessary condition
for such a starburst, as shown by Pompea \& Rieke (1990).

The strong association of nuclear and circumnuclear star formation
with bar structure, and the virtual absence of any other dependence
on morphological type, contrasts sharply with the behavior of the 
disk SFRs.  This implies that the evolution
of the circumnuclear region is largely decoupled from that of 
the disk at larger radii.  The strong distinctions between 
early-type and late-type barred galaxies
appear to be associated with the structural and dynamical properties
of the bars.  Bars in bulge-dominated, early-type spirals tend to be
very strong and efficient at transporting gas from the disk into the
central regions, while bars in late-type galaxies are much weaker
and are predicted to be much less efficient in transporting gas
(e.g. Athanassoula 1992, Friedli \& Benz 1995).  
All of the results are consistent with a general picture in which 
the circumnuclear SFRs of galaxies
are determined largely by the rate of gas transport into the 
nuclear regions.  

4.2.3~~GALAXY INTERACTIONS AND MERGERS~~~Numerous observations have
established a clear causal link between strong nuclear starbursts
and tidal interactions and mergers of galaxies.  Since this subject
is reviewed in depth elsewhere (Heckman 1990, 1994, Barnes \& Hernquist 1992,
Sanders \& Mirabel 1996, Kennicutt et al 1998), only the main results are
summarized here.

The evidence for interaction-induced nuclear star formation comes
from two types of studies, statistical comparisons of the SFRs in 
complete samples of interacting and non-interacting galaxies, and 
studies of the frequency of interactions and mergers among samples
of luminous starburst galaxies. 
Keel et al (1985) and Bushouse (1986) showed that 
the nuclear \halpha\ emission in nearby samples of interacting spiral galaxies 
is 3--4 times stronger than that in a control sample of isolated spirals.
Groundbased 10--20$\mu$m observations of the nuclear regions of 
interacting and merging galaxies showed similar or stronger enhancements,
depending on how the samples were selected 
(Lonsdale et al 1984, Cutri \& McAlary 1985, Joseph \& Wright 1985, Wright 
et al 1988).  There is an enormous range of SFRs among individual objects.
Spatially resolved data also show a stronger central concentration
of the star formation in strongly interacting systems 
(e.g. Bushouse 1987, Kennicutt et al 1987, Wright et al 1988).
Thus while the interactions tend to increase the SFR throughout galaxies,
the effects in the nuclear regions are stronger.  This radial concentration
is consistent with the predictions of numerical simulations of interacting
and merging systems 
(Barnes \& Hernquist 1992, Mihos \& Hernquist 1996, Kennicutt et al 1998).

The complementary approach is to measure the frequencies of interacting
systems in samples of starburst galaxies.
The most complete data of this kind come from IRAS, and they 
show that the importance of tidal triggering is a strong function
of the strength of the starburst,  
with the fraction of interactions increasing from 20--30\%\ for 
$L_{IR}$$<$10$^{10}~L_\odot$ to 70--95\%\ for $L_{IR}$$>$10$^{12}~L_\odot$. 
The relatively low fraction (Sanders et al 1988, Lawrence et al 
1989, Gallimore \& Keel 1993, Leech et al 1994, Sanders \& Mirabel 1996).
The relatively low fraction 
for the lower luminosity starbursts is understandable, because 
the corresponding SFRs ($<$1~\sfr) can be sustained with relatively
modest gas supplies, and can be fed by internal mechanisms such as a 
strong bar.  The most luminous starbursts, on the other hand, are 
associated almost exclusively with strong tidal interactions and mergers.
SFRs larger than $\sim$20~\sfr\ are rarely observed in isolated galaxies,
though some possible exceptions have been identified by Leech et al (1994).
In view of the enormous fueling requirements for such starbursts 
(Equations 5 and 6), however, it is perhaps not surprising that 
an event as violent as a merger is required.  These results underscore
the heterogeneous nature of the starburst galaxy population,
and they suggest that several triggering mechanisms are involved
in producing the population.

\section{INTERPRETATION AND IMPLICATIONS \\ FOR GALAXY EVOLUTION}

The observations described above can be fitted together into
a coherent evolutionary picture of disk galaxies and the Hubble
sequence.  This section summarizes the evolutionary implications 
of these data, taking into account the distinct patterns seen in
the disks and galactic nuclei.  It concludes with a discussion
of the critical role of the interstellar gas supply in regulating 
the SFR, across the entire range of galaxy types and environments.

\subsection{{\it Disk Evolution Along the Hubble Sequence}}

The strong trends observed in the SFR per unit luminosity along the
Hubble sequence mirror underlying trends in ther past star formation
histories of disks (Roberts 1963, Kennicutt 1983a, Gallagher et al 1984, 
Sandage 1986, Kennicutt et al 1994).  A useful parameter for characterizing
the star formation histories is the ratio of the current SFR to the
past SFR averaged over the age of the disk, denoted $b$ by Scalo (1986).
The evolutionary synthesis models discussed in Section 2 provide relations
between $b$ and the broadband colors and \halpha\ EWs.  Figure 3 
shows the distribution of $b$ (right axis scale) for an \halpha-selected 
sample of galaxies, based on the calibration of Kennicutt et al (1994).  
The typical late-type spiral has formed stars at
a roughly constant rate ($b \sim 1$), which is consistent with direct
measurements of the stellar age distribution in the Galactic disk
(e.g., Scalo 1986).  By contrast, early-type spiral galaxies are characterized
by rapidly declining SFRs, with $b \sim 0.01 - 0.1$, whereas elliptical
and S0 galaxies have essentially ceased forming stars ($b = 0$).  
Although the values of $b$ given above are based solely on synthesis
modelling of the \halpha\ equivalent widths, analysis of the integrated
colors and spectra of disks yield similar results (e.g. Kennicutt 1983a,
Gallagher et al 1984, Bruzual \& Charlot 1993, Kennicutt et al 1994).

The trends in $b$ shown in Figure 3 are based on integrated measurements,
so they are affected by bulge and nuclear contamination, which bias
the trends seen along the Hubble sequence.  A more detailed analysis
by Kennicutt et al (1994) includes corrections for bulge contamination
on the \halpha\ EWs.  The mean value of $b$ (for the disks alone) increases
from $<$0.07 for Sa disks, to 0.3 for Sb disks and 1.0 for Sc disks.
This change is much larger than the change in bulge mass fraction over
the same range of galaxy types, implying that most of the variation in
the integrated photometric properties of spiral galaxies is produced by changes
in the star formation histories of the disks, not in the bulge-to-disk
ratio.  Variations in bulge-disk structure may be play an important role,
however, in physically driving the evolution of the disks.

As discussed earlier, this picture has been challenged by 
Devereux \& Hameed (1997), based on the much weaker variation in 
FIR luminosities along the Hubble sequence.  The results
of the previous section provide part of the resolution to this paradox.
Many early-type barred spirals harbor luminous
circumnuclear starbursts, with integrated SFRs that can be as high
as the disk SFRs in late-type galaxies.  If this nuclear star formation is
included, then the interpretation of the Hubble sequence 
given above is clearly oversimplistic.  For that reason it is important
to delineate between the the nuclear regions and more extended disks
when characterizing the evolutionary properties of galaxies.  Much of the
remaining difference in interpretations hinges on the nature of the 
FIR emission in early-type galaxies, which may not directly trace the SFR
in all galaxies.

Although Figure 3 shows a strong change in the {\it average} star formation
history with galaxy type, it also shows a large dispersion in $b$
among galaxies of the same type.  Some of this must be due to 
real long-term variations in star formation history, reflecting 
the crudeness of the Hubble classification itself.  Similar ranges
are seen in the gas contents (Roberts \& Haynes 1994), and these 
correlate roughly with the SFR and $b$ variations (Figure 5).
Short-term variations in the SFR can also explain part of the 
dispersion in $b$.  Nuclear starbursts clearly play a role in some  
galaxies, especially early-type barred galaxies, and interaction-induced
starbursts are observed in a small percentage of nearby galaxies.  Starbursts  
are thought to be an important, if not the dominant mode of
star formation in low-mass galaxies (e.g. Hunter \& Gallagher 1985,
Hodge 1989, Ellis 1997), but the role of  
large-scale starbursts in massive galaxies is less well established,
(Bothun 1990, Kennicutt et al 1994, Tomita et al 1996).
A definitive answer to this question will probably come from 
lookback studies of large samples of disk galaxies.

\begin{figure}[!ht]
  \begin{center}
    \leavevmode
  \centerline{\epsfig{file=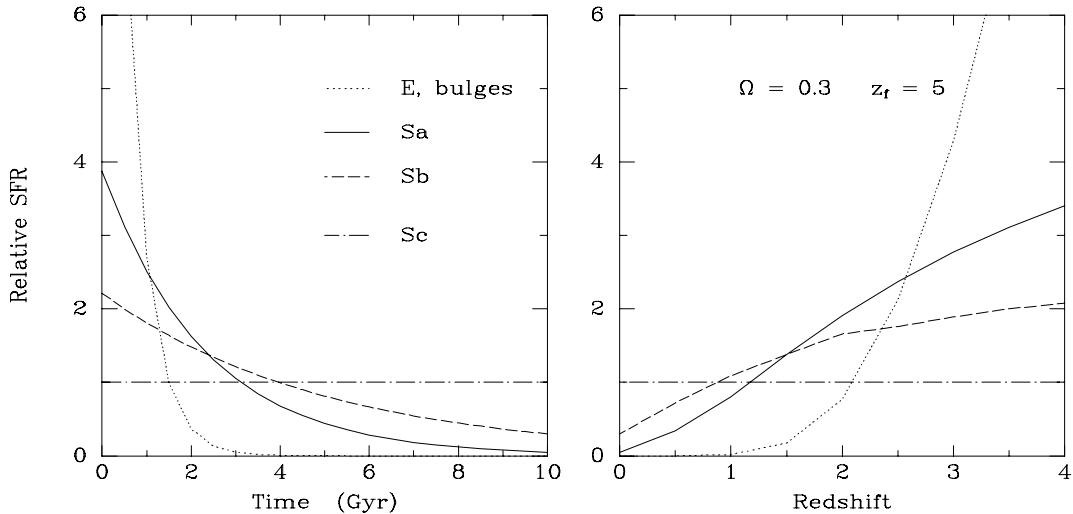,width=16cm}}
  \end{center}
  \caption{\em A schematic illustration of the evolution of the stellar 
birthrate for different Hubble types.  The left panel shows the
evolution of the relative SFR with time, following Sandage (1986).
The curves for spiral galaxies are exponentially declining SFRs that
fit the mean values of the birthrate parameter $b$ measured by Kennicutt
et al (1994).  The curve for elliptical galaxies and bulges is an arbitrary
dependence for a e-folding time of 0.5 Gyr, for comparative purposes only.
The right panel shows the corresponding evolution in SFR with redshift,
for an assumed cosmological density parameter $\Omega = 0.3$ and an
assumed formation redshift $z_f = 5$.}
\end{figure}

A schematic illustration of the trends in star formation histories
is shown in Figure 8.  The left panel compares the stellar birthrate
histories of typical elliptical galaxies (and spiral bulges), and the
disks of Sa, Sb, and Sc galaxies, following Sandage (1986).  
The curves for the spiral disks are exponential functions which correspond
to the average values of $b$ from Kennicutt et al (1994).  For illustrative
purposes, an exponentially declining SFR with an e-folding timescale of
0.5 Gyr is also shown, as might be appropriate for an old spheroidal 
population.  In this simple picture the Hubble sequence is primarily
one dictated by the characteristic timescale for star formation.
In the more contemporary hierarchical pictures of galaxy formation, 
these smooth histories would be punctuated by merger-induced starbursts,
but the basic long-term histories would be similar, especially for the disks.

The righthand panel in Figure 8 shows the same star formation histories,
but transformed into SFRs as functions of redshift (assuming $\Omega$=0.3 and
a formation redshift $z_f$=5).  This diagram illustrates how the dominant
star forming (massive) host galaxy populations might evolve with redshift.
Most star formation at the present epoch resides in late-type gas-rich
galaxies, but by $z\sim1$ all spiral types are predicted to have comparable 
SFRs, and
(present-day) early-type systems become increasingly dominant at higher
redshifts.  The tendency of early-type galaxies to have higher masses
will make the change in population with redshift even stronger. It will
interesting to see whether these trends are observed in future lookback
studies.  Many readers are probably aware that the redshift dependence of the
volume averaged SFR shows quite a different character, with a broad
maximum between $z$$\sim$1--2 and a decline at higher redshifts (Madau
et al 1996, 1998).  This difference probably reflects the importance
of hierarchical processes such as mergers in the evolution of galaxies,
mechanisms which are not included in the simple phenomenological description
in Figure 8 (Pei \& Fall 1995, Madau et al 1998).  

\subsection{{\it Evolution of Circumnuclear Star Formation}}

The SFRs in the circumnuclear regions are largely decoupled from
those of disks, and show no strong relationship to either the gas
contents or the bulge/disk properties of the parent galaxies.  
Instead, the nuclear SFRs are closely associated with dynamical
influences such as gas transport by bars or external gravitational
perturbations, which stimulate the flow of gas into the circumnuclear regions.

The temporal properties of the star formation in the nuclear
regions show a wide variation.  
Approximately 80--90\%\ of spiral nuclei in optically-selected samples
exhibit modest levels of Balmer emission, 
with an average \halpha\ emission-line equivalent width
of 20--30 \AA\ (Stauffer 1982, Kennicutt et al 1989b, Ho et al 1997a, b).
This is comparable to the average value in the disks of late-type
spiral galaxies, and is in the range expected for constant star 
formation over the age of the disk (Kennicutt 1983a, Kennicutt et al 1994).
Hence most nuclei show SFRs consistent with steady-state or declining
star formation, though it is likely 
that some of these nuclei are observed in a quiescent stage between 
major outbursts.

Starbursts are clearly the dominant mode of star formation in
IR-selected samples of nuclei.  The typical gas consumption times are 
in the range $10^8 - 10^9$ years (Figure 7), so the high SFRs can only
be sustained for a small percentage of the Hubble time.  These timescales
can be extended if a steady supply is introduced from the outside,
for example by a strong dissipative bar.  The most luminous nuclear
starbursts ($L_{bol}$$\ge$10$^{12}~L_\odot$) are singular events. 
Maintaining such luminosities for even $10^8$ years requires a total gas mass
on the order of $10^{10}$--$10^{11}~M_\odot$, equivalent to the total gas 
gas supply in most galaxies.  Violent interactions 
and mergers are the only events capable of triggering such a 
catastrophic mass transfer.

\subsection{{\it Physical Regulation of the SFR}}

Although star forming galaxies span millionfold ranges in their present SFRs
and physical conditions, there is a remarkable continuity in some of
the their properties, and these relationships provide 
important insights into the physical regulation of the SFR over this
entire spectrum of activities. 

\begin{figure}[!ht]
  \begin{center}
    \leavevmode
  \centerline{\epsfig{file=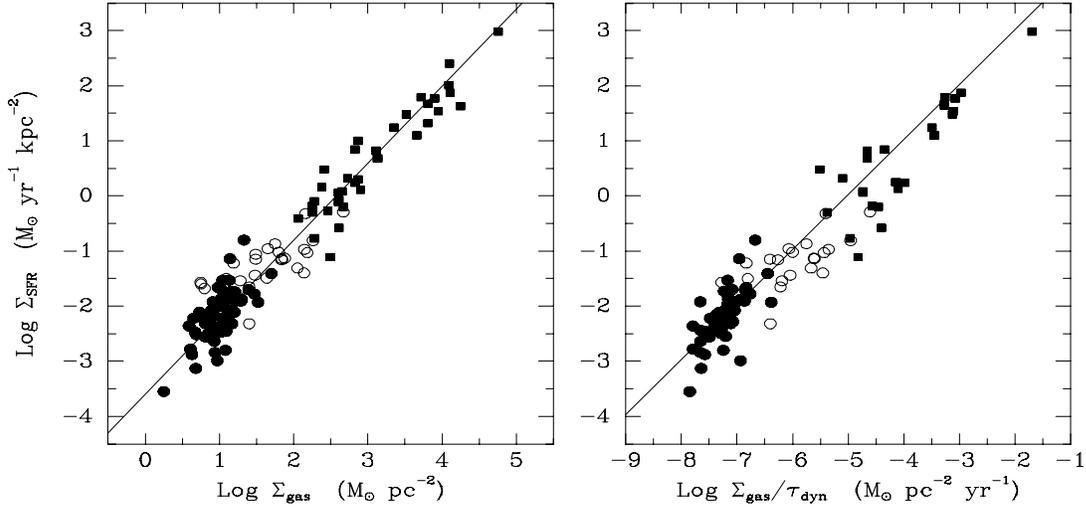,width=16cm}}
  \end{center}
  \caption{\em (Left) The global Schmidt law in galaxies. Solid points denote
the normal spirals in Figure 5, squares denote the circumnuclear starbursts
in Figure 7.  The open circles show the SFRs and gas densities of the
central regions of the normal disks. (Right) The same SFR data, but plotted
against the ratio of the gas density to the average orbital time in the 
disk.  Both plots are adapted from Kennicutt (1998).}
\end{figure}

We have already seen evidence from Figures 5 and 7 that the global 
SFRs of disks and nuclear starbursts are correlated
with the local gas density, though over very different ranges in
density and SFR per unit area.  The left panel of Figure 9 shows both sets 
of data plotted on a common scale, and it reveals that the entire range
of activities, spanning 5--6 orders of magnitude in gas and SFR densities,
fit on a common power law with index $N$$\sim$1.4 
(Kennicutt 1998).  The SFRs for the two sets of data were derived using
separate methods (\halpha\ luminosities for the normal disks and FIR 
luminosities for the starbursts), and to verify that they are measured  
on a self-consistent scale, Figure 9 also shows \halpha-derived SFRs
gas densities for the centers (1--2 kpc) of the normal disks (plotted
as open circles).  The tight relation shows that a simple Schmidt (1959)
power law provides an excellent empirical parametrization of the SFR,
across an enormous range of SFRs, and it suggests that the gas density
is the primary determinant of the SFR on these scales.

The uncertainty in the slope of the best fitting Schmidt law is dominated
by systematic errors in the SFRs, with the largest being the FIR-derived
SFRs and CO-derived gas densities
in the starburst galaxies.  Changing either scale individually by a
factor of two introduces a change of 0.1 in the fitted value of $N$,
and this is a reasonable estimate of the systematic errors involved
(Kennicutt 1998).  Incorporating these uncertainties yields the 
following relation for the best-fitting Schmidt law:
\begin{equation}
\Sigma_{SFR} = {{{(2.5 \pm 0.7)} \times 10^{-4}}~{({\Sigma_{gas} \over 
{1~M_\odot~{\rm pc}^{-2}}})^{1.4\pm0.15}}~M_\odot~{\rm yr^{-1}~kpc^{-2}}}.
\end{equation}
\noindent
where $\Sigma_{SFR}$ and $\Sigma_{gas}$ are the disk-averaged SFR and
gas densities, respectively.

As discussed by Larson (1992) and Elmegreen (1994), 
a large-scale Schmidt law with index $N$$\sim$1.5 would be expected 
for self-gravitating disks, if the SFR scales as the ratio of the gas
density ($\rho$) to the free-fall timescale ($\propto \rho^{-0.5}$),
and the average gas scale height is roughly constant across the sample
($\Sigma \propto \rho$).  In a variant on this picture, Elmegreen (1997) and
Silk (1997)
have suggested that the SFR might scale with the ratio of the 
gas density to the average orbital timescale; this is equivalent to
postulating that disks process a fixed fraction of their gas into
stars in each orbit around the galactic center.  The right panel
of Figure 9, also adapted from Kennicutt (1998), shows the correlation
between the SFR density and $\Sigma_{gas}/\tau_{dyn}$ for the same
galaxies and starbursts.  For this purpose $\tau_{dyn}$ is defined
as one disk orbit time, measured at half of the outer radius of the
star forming disk, in units of years (see Kennicutt 1998 for details).
The line is a median fit to the normal
disks with slope contrained to unity, as predicted by the simple
Silk model.  This alternative ``recipe" for the SFR provides a fit
that is nearly as good as the Schmidt law.  The equation of the 
fit is:  
\begin{equation}
\Sigma_{SFR} = 0.017~\Sigma_g~\Omega_g .
\end{equation}

In this parametrization the SFR is simply $\sim$10\%\ of the available 
gas mass per orbit.

These parametrizations offer two distinct interpretations of the high
SFRs in the centers of luminous
starburst galaxies.  In the context of the Schmidt law picture,
the star formation efficiency scales as $\Sigma{_g^{(N-1)}}$, or 
$\Sigma{_g^{0.4}}$ for the observed relation in Figure 9.  
The central starbursts have densities that are on the order of 100--10000
times higher than in the extended star forming disks of spirals,
so the global star formation efficiencies should be 6--40 times higher.
Alternatively, in the kinematical picture, the higher efficiencies in the
circumnuclear starbursts are simply a consequence of the shorter
orbital timescales in the galaxy centers, independent of the 
gas density.  Whether the observed Schmidt law is a consequence
of the shorter dynamical times or {\it vice versa} cannot be ascertained
from these data alone, but either description provides an excellent
empirical description or ``recipe" for the observed SFRs.

These simple pictures can account for the high SFRs in 
the starburst galaxies, as well as for the observed radial variation
of SFRs within star forming disks (Kennicutt 1989, 1997).  However
the relatively shallow $N$$\sim$1.4 Schmidt law cannot account for 
the strong changes in disk SFRs observed across the Hubble sequence,
if the disks evolved as nearly closed systems (Kennicutt et al 1994).  
Likewise the modest changes in galaxy rotation curves with Hubble type
are too small to account for the large differences in star formation 
histories with a kinematical model such as in equation [8].  The explanation 
probably involves star formation thresholds in the gas-poor galaxies
(Kennicutt 1989, Kennicutt et al 1997),  
but the scenario has not been tested quantitatively, and it is possible
that other mechanisms, such as infall of gas, merger history, or
bulge-disk interactions are responsible for the strong changes in
star formation histories across the spiral sequence.

\section{FUTURE PROSPECTS}

The observations described in this review have provided us with the
beginnings of a quantitative picture of the star formation properties
and evolutionary properties of the Hubble sequence.  However the
picture remains primitive in many respects, being based in large
part on integrated, one-zone averages over entire galaxies,
and extrapolations from present-day SFRs to crude characterizations of 
the past star formation histories.  Uncertainties in fundamental 
parameters such as the IMF and massive stellar evolution undermine the 
accuracy of the entire SFR scale, and weaken the interpretations that
are based on these measurements.  Ongoing work on several fronts should
lead to dramatic progress over the next decade, however.

The most exciting current development is the application of the 
SFR diagnostics described in Section 2 to galaxies spanning the full
range of redshifts and lookback times (Ellis 1997).  This work has
already provided the first crude measures of the evolution in the 
volume-averaged SFR (Madau et al 1996, 1998).  The combination of 
8--10 meter class groundbased telescopes,
HST, and eventually the {\it Next Generation Space Telescope} should
eventually provide detailed inventories of integrated spectra, SFRs and 
morphologies for complete samples of galaxies at successive redshifts.
This should give the definitive picture of the star formation history
of the Hubble sequence, and impose strong tests on galaxy formation
and evolution models.  At the same time, a new generation of IR
space observatories, including the {\it Wide Field Infrared Explorer} and
the {\it Space Infrared Telsescope Facility},
will provide high-resolution observations of nearby starburst galaxies,
and the first definitive measurements of the cosmological evolution
of the infrared-luminous starburst galaxy population.

Although studies of the star formation histories of nearby galaxies
are largely being supplanted by the more powerful lookback studies,
observations of nearby galaxies will remain crucial for understanding 
many critical aspects of galaxy formation and evolution.  Perhaps the greatest
potential is for understanding the physical processes that determine
the local and global SFRs in galaxies, and understanding the feedback
processes between the star formation and the parent galaxies.  This requires
spatially-resolved measurements of SFRs over the full
spectrum of insterstellar and star formation environments, and complementary
measurements of the densities, dynamics, and abundances of the 
interstellar gas.  Uncertainty about the nature of the star formation
law and the SFR--ISM feedback cycle remain major stumbling
blocks to realistic galaxy evolution models, but observations over
the next decade should provide the foundations of a 
physically-based model of galactic star formation and the Hubble sequence.

\section{ACKNOWLEDGEMENTS}

I wish to express special thanks to my collaborators in the research
presented here, especially my current and former graduate students
Audra Baleisis, Fabio Bresolin, Charles Congdon, Murray Dixson, 
Kevin Edgar, Paul Harding, Crystal Martin, Sally Oey, Anne Turner,
and Rene Walterbos.  During the preparation of this review
my research was supported by the National 
Science Foundation though grant AST-9421145.

\newpage

{\it Literature Cited}

\medskip
\parskip=5pt

Aalto S, Booth RS, Black JH, Koribalski B., Wielebinski R.
     1994. {\it Astron. Astrohys.} 286:365-80

Athanassoula E. 1992. {\it MNRAS} 259:345-64

Bagnuolo WG. 1976. {\it The Stellar Content and Evolution of Irregular
 and Other Late-Type Galaxies}.  PhD thesis.  Caltech

Balzano VA. 1983. {\it Ap. J.} 268:602-27

Barnes JE, Hernquist L. 1992. {\it Annu. Rev. Astron. Astrophys.} 30:705-42

Bechtold J, Yee HKC, Elston R, Ellingson E. 1997. {\it Ap. J. Lett.} 
  477:L29-L32

Bertelli G, Bressan A, Chiosi C, Fagotto F, Nasi E. 1994. {\it
  Astron. Astrophys. Suppl.} 106:275-302

Boselli A. 1994. {\it Astron. Astrophys.} 292:1-12

Boselli A, Gavazzi G, Lequeux J, Buat V, Casoli F, et al. 1995.
{\it Astron. Astrophys.} 300:L13-L16

Bothun GD. 1990. In {\it Evolution of the Universe of Galaxies,}
  ed. RG Kron, {ASP Conf. Proc.} 10:54-66, San Francisco: Astron. Soc. Pac.

Bresolin F, Kennicutt RC. 1997, {\it Astron. J.} 113:975-80

Bruzual G, Charlot S. 1993. {\it Ap. J.} 405:538-53

Buat V. 1992. {\it Astron. Astrophys.} 264:444-54

Buat V, Deharveng JM. 1988. {\it Astron. Astrophys.} 195:60-70

Buat V, Deharveng JM, Donas J. 1989. {\it Astron. Astrophys.} 223:42-46

Buat V, Xu C. 1996. {\it Astron. Astrophys.} 306:61-72

Bushouse HA. 1986. {\it Astron. J.} 91:255-70

Bushouse HA. 1987. {\it Ap. J.} 320:49-72

Bushouse HA, Werner MW, Lamb SA. 1988. {\it Ap. J.} 335:74-92

Caldwell N, Kennicutt R, Phillips AC, Schommer RA. 1991. {\it Ap. J.}
370:526-40

Caldwell N, Kennicutt R, Schommer R. 1994. {\it Astron.J. 108:1186-90}

Calzetti D, Kinney AL, Storchi-Bergmann T. 1994. {\it Ap.J.} 429:582-601

Calzetti D, Kinney AL, Storchi-Bergmann T. 1996. {\it Ap. J.} 458:132-5

Calzetti D. 1997. {\it Astron. J.} 113:162-84

Caplan J, Deharveng L. 1986. {\it Astron. Astrophys.} 155:297-313

Caplan J, Ye T, Deharveng L, Turtle AJ, Kennicutt RC. 1996. {\it Astron.
 Astrophys.} 307:403-16

Carico DP, Sanders DB, Soifer BT, Matthews K, Neugebauer G. 1990. 
  {\it Astron. J.} 100:70-83

Cayatte V, Kotanyi C, Balkowski C, van Gorkom JH. 1994. {\it Astron. J.}
   107:1003-17

Cepa J, Beckman JE 1990. {\it Ap. J.} 349:497-502

Charlot S, Bruzual G. 1991. {\it Ap. J.} 367:126-40

Cohen JG. 1976. {\it Ap. J.} 203:587-92

Cox, P, Mezger PG. 1989. {\it Astron. Astrophys. Rev.} 1:49-83

Cowie LL, Hu EM, Songaila A, Egami E. 1997. {\it Ap. J. Lett.} 481:L9-L13

Cowie LL, Songaila A, Hu EM, Cohen JG. 1996. {\it Astron. J.} 112:839-64

Cutri RM, McAlary CW. 1985. {\it Ap. J.} 296:90-105

Deharveng JM, Sasseen TP, Buat V, Bowyer S, Lampton M, 
  Wu X. 1994. {\it Astron. Astrophys.} 289:715-728

de Vaucouleurs G, de Vaucouleurs A, Corwin HG. 1976. {\it Second Reference
 Catalog of Bright Galaxies}.  Austin: Univ. of Texas Press (RC2)

Devereux N. 1987. {\it Ap. J.} 323:91-107

Devereux NA, Becklin EE, Scoville N. 1987. {\it Ap. J.} 312:529-41

Devereux NA, Hameed S. 1997. {\it Astron. J.} 113:599-608

Devereux NA, Young JS. 1990. {\it Ap. J. Lett.} 350:L25-28

Devereux NA, Young JS. 1991. {\it Ap. J.} 371:515-24

Donas J, Deharveng JM. 1984. {\it Astron. Astrophys.} 140:325-333

Donas J, Deharveng JM, Laget M, Milliard B, Huguenin D. 1987.
{\it Astron. Astrophys.} 180:12-26 

Donas J, Milliard B, Laget M, Buat V.  1990. {\it Astron. Astrophys.}
  235:60-68

Donas J, Milliard B, Laget M. 1995. {\it Astron. Astrophys.} 303:661-672

Downes D, Solomon PM, Radford SJE. 1993. {\it Ap. J. Lett.} 414:L13-L16

Dressel LL. 1988. {\it Ap. J. Lett.} 329:L69-L73

Dressler A. 1984. {\it Annu. Rev. Astron. Astrophys.} 22:185-222

Ellis RS. 1997. {\it Annu. Rev. Astron. Astrophys.} 35:389-443

Elmegreen BG. 1994. {\it Ap. J. Lett.} 425:L73-76

Elmegreen BG. 1997, In {\it Starburst Activity in Galaxies}, ed J Franco,
R Terlevich, A Serrano, {\it Rev. Mex. Astron. Astrophys. Conf. Ser.}
6:165-71

Elmegreen BG, Elmegreen DM. 1986. {\it Ap. J.} 311:554-562

Engelbracht CW. 1997. {\it Infrared Observations and Stellar Populations
Modelling of Starburst Galaxies.}  PhD thesis, Univ. Arizona

Evans IN, Koratkar AP, Storchi-Bergmann T, Kirkpatrick H, Heckman TM,
Wilson AS. 1996. {\it Ap. J. Suppl.} 105:93-127

Fanelli MN, Marcum PM, Waller WH, Cornett RH, O'Connell RW, et al. 1997.
In {\it The Ultraviolet Universe at Low and High Redshift,} ed. W Waller,
M Fanelli, J Hollis, A Danks.  New York: Am. Inst. Phys. 

Feinstein C. 1997. {\it Ap. J. Suppl.} 112:29-47

Ferguson AMN, Wyse RFG, Gallagher JS, Hunter DA. 1996. {\it Astron. J.}
  111:2265-79

Fioc M, Rocca-Volmerange B. 1997. {\it Astron. Astrophys.} 326:950-62

Friedli D, Benz W. 1995. {\it Astron. Astrophys.} 301:649-65

Gallagher JS, Hunter DA. 1984. {\it Ann. Rev. Astron. Astrophys.} 22:37-74

Gallagher JS, Hunter DA, Bushouse H. 1989. {\it Astron. J.} 97:700-07

Gallagher JS, Hunter DA, Tutukov AV. 1984. {\it Ap. J.} 284:544-56

Gallego J, Zamorano J, Aragon-Salamanca A, Rego M. 1995. {\it Ap. J. Lett.}
  445:L1-L4

Gallimore JF, Keel WC. 1993. {\it Astron. J.} 106:1337-43

Gavazzi G, Boselli A, Kennicutt R. 1991. {\it Astron. J.} 101:1207-30

Gavazzi G, Jaffe W. 1985. {\it Ap. J. Lett.} 294:L89-L92

Gavazzi G, Pierini D, Boselli A. 1996. {\it Astron. Astrophys.} 312:397-408

Gavazzi G, Scodeggio M. 1996. {\it Astron. Astrophys.} 312:L29-L32

Gilmore GF, Howell DJ. (eds.) 1998. {\it The Stellar Initial Mass
Function.}, {\it ASP Conf. Proc.}, Vol. 142. San Francisco: Astron. Soc. Pac.

Giuricin G, Tamburini L, Mardirossian F, Mezzetti M, Monaco P. 1994.
{\it Astron. Astrophys.} 427:202-20

Goldader JD, Joseph RD, Doyon R, Sanders DB. 1995. {\it Ap. J.} 444:97-112

Goldader JD, Joseph RD, Doyon R, Sanders DB. 1997. {\it Ap. J. Suppl.}
 108:449-70

Gonz\'alez Delgado, RM, Perez E, Tadhunter C, Vilchez J, Rodr\'iguez-Espinoza
 JM. 1997. {\it Ap. J. Suppl.} 108:155-98

Hawarden TG, Mountain CM, Leggett SK, Puxley PJ. 1986. {\it MNRAS} 221:41P-45P

Harper DA, Low FJ. 1973. {\it Ap. J. Lett.} 182:L89-L93

Haynes MP, Giovanelli R, Chincarini GL. 1984. {\it Ann. Rev. Astron. 
  Astrophys.} 22:445-70

Heckman TM 1990.  In {\it Paired and Interacting Galaxies, IAU Colloq. 124,}
  ed. JW Sulentic, WC Keel, CM Telesco, NASA 
  Conf. Publ. CP-3098, pp. 359-82.  Washington DC: NASA

Heckman TM. 1994.  In 
{\it Mass-Transfer Induced Activity in Galaxies}, ed. I Shlosman,
pp. 234-50.  Cambridge: Cambridge Univ. Press

Heckman TM, Crane PC, Balick B. 1980. {\it Astron. Astrophys. Suppl.}
  40:295-305

Heiles C. 1990. {\it Ap. J.} 354:483-91

Ho LC, Filippenko AV, Sargent WLW. 1997. {\it Ap. J.} 487:579-90

Ho LC, Filippenko AV, Sargent WLW. 1997. {\it Ap. J.} 487:591-602

Ho PTP, Beck SC, Turner JL. 1990. {\it Ap. J.} 349:57-66

Hodge PW. 1989. {\it Annu. Rev. Astron. Astrophys.} 27:139-59

Hodge PW, Kennicutt RC. 1983. {\it Astron. J.} 88:296-328

Huang JH, Gu QS, Su HJ, Hawarden TG, Liao XH, Wu GX. 1996: {\it Astron.
  Astrophys.} 313:13-24

Hubble E. 1926. {\it Ap. J.} 64:321-69

Huchra JP. 1977. {\it Ap. J.} 217:928-39

Hunter DA. 1994. {\it Astron. J.} 107:565-81

Hunter DA, Gallagher JS. 1985. {\it Ap. J. Suppl.} 58:533-60

Hunter DA, Gallagher JS. 1990. {\it Ap. J.} 362:480-90

Hunter DA, Gillett FC, Gallagher JS, Rice WL, Low FJ. 1986.
 {\it Ap. J.} 303:171-85

Hunter DA, Hawley WN, Gallagher JS. 1993. {\it Astron. J.} 106:1797-1811

Isobe T, Feigelson E. 1992. {\it Ap. J. Suppl.} 79:197-211

Israel FP, van der Hulst JM. 1983. {\it Astron. J.} 88:1736-48

Joseph RD, Wright GS. 1985. {\it MNRAS} 214:87-95

Kaufman M, Bash FN, Kennicutt RC, Hodge PW. 1987. {\it Ap. J.} 319:61-75

Keel WC. 1983. {\it Ap. J.} 269:466-86

Keel WC, Kennicutt RC, Hummel E, van der Hulst JM. 1985. {\it Astron. J.}
  90:708-30

Kennicutt RC. 1983a. {\it Ap. J.} 272:54-67 

Kennicutt RC. 1983b. {\it Astron. J.} 88:483-88

Kennicutt RC. 1989. {\it Ap. J.} 344:685-703

Kennicutt RC. 1992a. {\it Ap. J.} 388:310-27

Kennicutt RC. 1992b. {\it Ap. J. Suppl.} 79:255-84

Kennicutt RC. 1997, In {\it The Interstellar Medium in Galaxies,}
 ed. JM van der Hulst, pp. 171-95.  Dordrecht: Kluwer

Kennicutt RC. 1998. {\it Ap. J.} 498:541-52

Kennicutt RC, Schweizer F, Barnes JE. 1998.  {\it Galaxies: Interactions
and Induced Star Formation, Saas-Fee Advanced Course 26,} ed. D Friedli,
L Martinet, D Pfenniger, Berlin:Springer

Kennicutt RC, Bothun GD, Schommer RA. 1984. {\it Astron. J.} 89:1279-87

Kennicutt RC, Bresolin F, Bomans DJ, Bothun GD, Thompson IB. 1995.
  {\it Astron. J.} 109:594-604

Kennicutt RC, Edgar BK, Hodge PW. 1989a. {\it Ap. J.} 337:761-81

Kennicutt RC, Keel WC, Blaha CA. 1989b. {\it Astron. J.} 97:1022-35

Kennicutt RC, Keel WC, van der Hulst JM, Hummel E, Roettiger KA.
1987.  {\it Astron. J.} 93:1011-23

Kennicutt RC, Kent SM. 1983. {\it Astron. J.} 88:1094-1107

Kennicutt RC, Tamblyn P, Congdon CW. 1994. {\it Ap. J.} 435:22-36

Kinney AL, Bohlin RC, Calzetti D, Panagia N, Wyse RFG. 1993. {\it 
Ap. J. Suppl.} 86:5-93

Klein U, Grave R. 1986. {\it Astron. Astrophys.} 161:155-68

Knapen J, Beckman JE, Cepa J, van der Hulst JM, Rand RJ. 1992. 
  {\it Ap. J. Lett.} 385:L37-L40

Larson RB. 1992, In {\it Star Formation in Stellar Systems},
 ed. G Tenorio-Tagle, M Prieto, F S\'anchez. pp. 125-190. 
 Cambridge: Cambridge Univ. Press

Larson RB, Tinsley BM. 1978. {\it Ap. J.} 219:46-59

Lawrence A, Rowan-Robinson M, Leech K, Jones DHP, Wall JV. 1989.
  {\it MNRAS} 240:329-48

Leech M, Rowan-Robinson M, Lawrence A, Hughes JD. 1994. {\it MNRAS} 267:253-69 

Lehnert MD, Heckman TM. 1996. {\it Ap. J.} 472:546-63

Leitherer C, Ferguson HC, Heckman TM, Lowenthal JD. 1995a. {\it Ap. J. Lett.}
  454:L19-L22

Leitherer C, Fritze-v. Alvensleben U, Huchra JP. (eds.) 1996b. {\it 
From Stars to Galaxies: The Impact of Stellar Physics on Galaxy
Evolution.} {\it ASP Conf. Proc.} Vol. 98. San Francisco: Astron. Soc Pac.

Leitherer C, Heckman TM. 1995. {\it Ap. J. Suppl.} 96:9-38

Leitherer C, Robert C, Heckman TM. 1995b. {\it Ap. J. Suppl.} 99:173-87

Leitherer C, Alloin D, Alvensleben UF, Gallagher JS, Huchra JP, et al.
  1996a. {\it Publ. Astron. Soc. Pac.} 108:996-1017

Liu CT, Kennicutt RC. 1995. {\it Ap. J.} 450:547-58

Lonsdale CJ, Helou G. 1987. {\it Ap. J.} 314:513-24

Lonsdale CJ, Persson SE, Matthews K. 1984. {\it Ap. J.} 287:95-107

Lutz D, Genzel R, Sternberg A, Netzer H, Kunze D, et al. 1996. 
  {\it Astron. Astrophys.} 315:L137-L140

Madau P, Ferguson H, Dickinson M, Giavalisco M, Steidel CC, Fruchter A.
 1996. {\it MNRAS} 283:1388-1404

Madau P, Pozzetti L, Dickinson M. 1998. {\it Ap. J.} in press

Maoz D, Filippenko AV, Ho LC, Macchetto D, Rix H-W,  
  Schneider DP. 1996. {\it Ap. J. Suppl.} 107:215-26

Martin CL. 1997. {\it Ap. J.} in press

Martinet L, Friedli D. 1997. {\it Astron. Astrophys.} 323:363-73

Massey P. 1998.  In {\it The Stellar Initial Mass Function.} ed.
BF Gilmore, DJ Howell. {\it ASP Conf. Proc.} 142:17-44. San Francisco: 
Astron. Soc. Pac.

McCall ML, Schmidt FH. 1986. {\it Ap. J.} 311:548-53

Meurer GR, Heckman TM, Leitherer C, Kinney A, Robert C,  
  Garnett DR. 1995. {\it Astron. J.} 110:2665-91

Meurer GR, Gerhardt R, Heckman TM, Lehnert MD, Leitherer C, 
  Lowenthal J. 1997, {\it Astron. J.} 114:54-68

Mihos JC, Hernquist L. 1996. {\it Ap. J.} 464:641-63

Morgan WW. 1958. {\it Publ. Astron. Soc. Pac.} 70:364-91

Moss C, Whittle M. 1993. {\it Ap. J. Lett.200} 407:L17-L20

Moshir M, Kopan, G, Conrow J, McCallon H, Hacking P, et al. 1992.
{\it Explanatory Supplement to the IRAS Faint Source
Survey, Version 2}, JPL D-10015 8/92, (Pasadena: JPL)

Niklas S, Klein U, Braine J, Wielebinski R. 1995. {\it Astron. Astrophys.
  Suppl.} 114:21-49

Niklas S, Klein U, Wielebinski R. 1997.  {\it Astron. Astrophys.} 322:19-28

Norman C, Ikeuchi S. 1989. {\it Ap. J.} 345:372-83

Oey MS, Kennicutt RC. 1997. {\it MNRAS} 291:827-32

Ostriker JP, Thuan TX. 1975. {\it Ap. J.} 202:353-64

Patel K, Wilson CD. 1995a. {\it Ap. J.} 451:607-15

Patel K, Wilson CD. 1995b. {\it Ap. J.} 453:162-72

Pei YC, Fall SM. 1995. {\it Ap. J.} 454:69-76

Phillips AC. 1993. {\it Star Formation in Barred Spiral Galaxies.} 
 PhD thesis, Univ. Washington, Seattle

Pogge RW, Eskridge PB. 1987. {\it Astron. J.} 93:291-300

Pogge RW, Eskridge PB. 1993. {\it Astron. J.} 106:1405-19

Pompea SM, Rieke GH. 1990. {\it Ap. J.} 356:416-29

Puxley PJ, Brand PWJL, Moore TJT, Mountain CM, Nakai N, Yamashita AT.
 1989. {\it Ap. J.} 345:163-68

Puxley PJ, Hawarden TG, Mountain CM. 1990. {\it Ap. J.} 364:77-86

Rieke GH, Lebofsky MJ. 1978. {\it Ap. J. Lett.} 220:L37-L41

Rieke GH, Lebofsky MJ. 1979. {\it Ann. Rev. Astron. Astrophys.} 17:477-511

Rieke, GH, Loken K, Rieke MJ, Tamblyn P. 1993. {\it Ap. J.} 412:99-110

Rieke GH, Low FJ. 1972.  {\it Ap. J. Lett.} 176:L95-L100

Roberts MS. 1963. {\it Ann. Rev. Astron. Astrophys.} 1:149-78

Roberts MS, Haynes MP. 1994. {\it Ann. Rev. Astron. Astrophys.} 32:115-52

Romanishin W. 1990. {\it Astron. J.} 100:373-76

Rowan-Robinson M, Crawford J. 1989. {\it MNRAS} 238:523-58

Rubin VC, Kenney JDP, Young JS. 1997. {\it Astron. J.} 113:1250-78

Ryder SD. 1993. {\it Massive Star Formation in Galactic Disks.}  PhD
 thesis. Australian National Univ.

Ryder SD, Dopita MA. 1993. {\it Ap. J. Suppl.} 88:415-21

Ryder SD, Dopita MA. 1994. {\it Ap. J.} 430:142-62

Salpeter EE. 1955. {\it Ap. J.} 121:161-67

Sandage A. 1986. {\it Astron. Astrophys.} 161:89-101

Sanders DB, Mirabel IF. 1996. {\it Ann. Rev. Astron. Astrophys.} 34:749-92

Sanders DB, Scoville NZ, Soifer BT. 1991. {\it Ap. J.} 370:158-71

Sanders DB, Soifer BT, Elias JH, Madore, BF, Matthews K, et al. 1988.
{\it Ap. J.} 325:74-91

Sauvage M, Thuan TX. 1992. {\it Ap. J. Lett.} 396:L69-L73

Sauvage M, Thuan TX. 1994. {\it Ap. J.} 429:153-71

Scalo JM. 1986. {\it Fund. Cos. Phys.} 11:1-278

Schmidt M. 1959. {\it Ap. J.} 129:243-58 

Scoville NZ, Becklin EE, Young JS, Capps RW. 1983. {\it Ap. J.} 271:512-23

Scoville NZ, Hibbard JE, Yun MS, van Gorkom JH. 1994.  In 
{\it Mass-Transfer Induced Activity in Galaxies}, ed. I Shlosman,
pp. 191-212.  Cambridge: Cambridge Univ. Press

Searle L, Sargent WLW, Bagnuolo WG. 1973. {\it Ap. J.} 179:427-38

S\'ersic JL, Pastoriza M. 1967. {\it Publ. Astron. Soc. Pac} 79:152-55

Silk J. 1997. {\it Ap. J.} 481:703-09

Smith AM, Cornett, RH. 1982, {\it Ap. J.} 261:1-11

Smith BJ, Harvey PM. 1996. {\it Ap. J.} 468:139-66

Smith EP, Pica AJ, Bohlin RC, Cornett RH, Fanelli MN. 1996. {\it 
Ap. J. Suppl.} 104:207-315

Soifer BT, Houck JR, Neugebauer G. 1987. {\it Ann. Rev. Astron. Astrophys.}
25:187-230

Solomon PM, Downes D, Radford SJE, Barrett JW. 1997. {\it Ap. J.} 478:144-61

Solomon PM, Sage LJ. 1988. {\it Ap. J.} 334:613-25

Stauffer JR. 1982. {\it Ap. J. Suppl.} 50:517-27

Steidel CC, Giavalisco M, Pettini M, Dickinson M, Adelberger KL.
1996. {\it Ap. J. Lett.} 462:L17-L21

Telesco CM. 1988. {\it Ann. Rev. Astron. Astrophys.} 26:343-76

Telesco CM, Dressel LL, Wolstencroft RD. 1993. {\it Ap. J.} 414:120-43

Telesco CM, Harper DA. 1980. {\it Ap. J.} 235:392-404

Telesco CM, Wolstencroft RD, Done C. 1988. {\it Ap. J.} 329:174-86

Thronson HA, Bally J, Hacking P. 1989. {\it Astron. J.} 97:363-74

Tinney CG, Scoville NZ, Sanders DB, Soifer BT 1990, {\it Ap. J.} 362:473-79

Tinsley BM. 1968. {\it Ap. J.} 151:547-65

Tinsley BM. 1972. {\it Astron. Astrophys.} 20:383-96

Tomita A, Tomita Y, Saito M. 1996. {\it Pub. Astron. Soc. J.} 48:285-303

Tully RB, Mould JR, Aaronson M. 1982. {\it Ap. J.} 257:527-37

Turner JL, Ho PTP. 1994. {\it Ap. J.} 421:122-39

van der Hulst JM, Kennicutt RC, Crane PC, Rots AH. 1988.
 {\it Astron. Astrophys.} 195:38-52

Veilleux S, Kim D-C, Sanders DB, Mazzarella JM, Soifer BT. 1995. 
 {\it Ap. J. Suppl.} 98:171-217

Waller W, Fanelli M, Danks A, Hollis J. 1997. {\it The Ultraviolet 
Universe at Low and High Redshift, AIP Conf.} 408. New York: 
Am. Inst. Phys.

Walterbos RAM, Braun R. 1994. {\it Ap. J.} 431:156-71

Walterbos RAM, Greenawalt B. 1996. {\it Ap. J.} 460:696-710

Warmels RH. 1988. {\it Astron. Astrophys. Suppl.} 72:427-47

Weedman DW, Feldman FR, Balzano VA, Ramsey LW, Sramek RA,
Wu C-C. 1981. {\it Ap. J.} 248:105-12

Whitford AE. 1975. in {\it Galaxies in the Universe,} ed. A Sandage,
  M Sandage, J Kristian, {\it Stars Stellar Syst. Compend.} 9:159-76.
  Chicago: Univ. Chicago Press

Wright GS, Joseph RD, Robertson NA, James PA, Meikle WPS. 1988. 
  {\it MNRAS} 233:1-23

Wyse RFG. 1983. {\it MNRAS} 199:1P-8P

Xu C, Sulentic JW. 1991. {\it Ap. J.} 374:407-30

Young JS, Allen L, Kenney JDP, Lesser A, Rownd B. 1996. {\it Astron. J.}
  112:1903-27

Young JS, Scoville NZ. 1991. {\it Ann. Rev. Astron. Astrophys.} 29:581-625

Young JS, Schloerb, FP, Kenney JDP, Lord SD. 1986. {\it Ap. J.} 304:443-58

\end{document}